\documentclass[epj]{svjour}
\usepackage{graphics}
\usepackage{epsfig}
\usepackage{url}
\usepackage{hyphenat}
\usepackage{amsmath}

\def\etmiss{\big\slash\hspace{-1.6ex}{E_\mathrm{T}}}

\usepackage{graphicx}
\usepackage{epstopdf}

\begin{document}
\title{Measurements of the Production, Decay and Properties of the Top
  Quark: A Review}
\author{Kevin Lannon\inst{1} \and Fabrizio Margaroli\inst{2} \and Chris Neu\inst{3}
}                     
\offprints{K. Lannon}          
\institute{University of Notre Dame, Notre Dame, IN 46556, USA \and Sapienza Universit\`{a} di Roma and INFN Roma1, 00185 Rome, Italy \and University of Virginia, Charlottesville, VA 22904, USA}
\date{Received: date / Revised version: date}
%
\abstract{ With the full Tevatron Run II and early LHC data samples,
  the opportunity for furthering our understanding of the properties
  of the top quark has never been more promising.  Although the
  current knowledge of the top quark comes largely from Tevatron
  measurements, the experiments at the LHC are poised to probe
  top-quark production and decay in unprecedented regimes.  Although
  no current top quark measurements conclusively contradict
  predictions from the standard model, the precision of most
  measurements remains statistically limited.  Additionally, some
  measurements, most notably $A_{FB}$ in top quark pair production,
  show tantalizing hints of beyond-the-Standard-Model dynamics.  The
  top quark sample is growing rapidly at the LHC, with initial results
  now public.  This review examines the current status of top quark
  measurements in the particular light of searching for evidence of
  new physics, either through direct searches for beyond the standard
  model phenomena or indirectly via precise measurements of standard
  model top quark properties.
} 
\maketitle
\section{Introduction}
\label{sec:Intro}
The observation of the top quark at the Tevatron in 1995 marked the end of the search for the isospin partner to the bottom quark, completing the three-generation structure of the quark sector in the Standard Model.  Yet, it also marked the beginning of the quest to understand why the top quark is so different from the other quarks.  The very feature that allowed the top quark to evade experimental detection for so long---its extraordinarily high mass---is also the cause of most of its unusual properties.  The 173.2~GeV/c$^2$ mass of the top quark\,\cite{TevatronElectroweakWorkingGroup:2011wr} makes it roughly forty times more massive than the next most massive quark, the bottom quark, and over 10\,000 times more massive than the up quark.  Perhaps even more striking is that, as a fundamental object believed to possess no internal structure, the top quark is more massive than the first seventy-four elements in the periodic table.

The top quark's large mass has phenomenological implications as well.  Foremost, because the top quark is more massive than the $W$ boson, top quark decays proceed rapidly through te electroweak interaction via an on-shell $W$ before the top quark has a chance to form a hadronic bound state.  This makes the top quark the only quark whose properties can be studied without the complications of disentangling hadronization effects.  Furthermore, because $|V_{tb}| \sim 1$ in the Standard Model, the top quark decays to a $W$ boson and a $b$ almost 100\% of the time, producing a signature quite distinct from the collimated hadronic jets that signal the production of lighter quarks.  Finally, as a result of its large mass, a significant fraction (roughly 70\% in the Standard Model) of the the $W$ bosons produced in top quark decays are longitudinally polarized.

Beyond the phenomenological implications, the large top quark mass raises deeper questions.  In the Standard Model, quark masses arise from the quarks' couplings to the Higgs boson.  The large top quark mass implies the top quark has a particularly large coupling to the Higgs boson.  In fact, the specific value of the top quark mass implies a Yukawa coupling to the Higgs very near unity.  From this perspective it might be more appropriate to ask why the rest of the quarks have such an unnaturally small coupling to the Higgs rather than asking why the top quark mass is so large.  Regardless of whether the Higgs mechanism provides the correct explanation for the electroweak symmetry breaking, the large top quark mass raises the possibility that the top quark may have some special connection to or play some special role in the mechanism of electroweak symmetry breaking.  Several alternative models to the Higgs picture of electroweak symmetry breaking, such as the top quark see-saw theory~\cite{topSeeSaw} or top-color-assisted technicolor~\cite{topTechnicolor} posit such a role for the top quark.

Therefore, the primary focus of top quark physics is the search for some evidence of physics beyond the Standard Model -- in particular new physics that would help to explain the top quark's singular differences in comparison with the rest of the quark sector.  Strategies in the search for new physics associated with the top quark fall broadly into two categories:  First there are the direct searches for new physics associated with top quark production or decay.  Examples include searching for new heavy resonances decaying into top quarks or searching for new particles produced in top quark decays.  An alternative strategy is to measure properties of the top quark predicted within the Standard Model, such as its production cross section, both inclusively and differentially, as well as its decay branching fractions, looking for deviations compared to the Standard Model predictions.  This review considers analyses of both types and summarizes the current experimental picture of top quark physics.

\subsection{Accelerators and the Top Quark}

It is impossible to relate the tale of the top quark without mentioning the two accelerators that to date are the only locations where top quarks have been produced in a laboratory setting.  The first accelerator to produce top quarks was the Tevatron located at Fermi National Accelerator Laboratory in Batavia, Illinois in the United States.  The Tevatron collides protons with antiprotons in a 6.28 km circumference ring.  When it began colliding beams operations in 1985, the center of mass energy was 1.8~TeV, and before the end of the first collider run, referred to as Run I, it had achieved instantaneous luminosities in excess of $1.1 \times 10^{31}$~cm$^{-2}$s$^{-1}$.  Nearly ten years after the start of Tevatron operations, the CDF\cite{Abulencia:2005ix,Acosta:2004yw,cdfSecVtx,Hill:2004qb} and D0~\cite{Abazov:2005pn} experiments jointly announced first observation of top quark pair production using 67~pb$^{-1}$ and 50~pb$^{-1}$ of integrated luminosity respectively~\cite{cdfTopObs,d0TopObs}.  Shortly after the top quark observation, the Tevatron shut down for upgrades both to the maximum beam energy and luminosity, and in 2001, the second Tevatron run, known as Run II, began, with a center of mass energy of 1.96 TeV.  Roughly eight years later, with a dataset corresponding to 3.2~fb$^{-1}$ at CDF and 2.1~fb$^{-1}$ at D0, both experiments announced observation of production of single top quarks through the electroweak interaction~\cite{cdfSingleTopObs,d0SingleTopObs}.  The long time delay and large difference in dataset size needed for the observation of single top quark production compared to top quark pair production is a testament to the experimental difficulty of extracting the small single top signal from the large backgrounds.  On Sep. 30,  2011, after 26 years of colliding beams, the Tevatron accelerator ceased operations.  Over that time period, the Tevatron delivered an integrated luminosity of almost 12~fb$^{-1}$ and achieved instantaneous luminosities as high as $4.1 \times 10^{32}$~cm$^{-2}$s$^{-1}$.  A wide range of top quark analysis continues to be pursued using this dataset both at CDF and D0.

In March of 2010, the Large Hadron Collider (LHC) located at the CERN laboratory in Geneva,  Switzerland, significantly extended the energy frontier, reaching a center of mass energy of 7~TeV.  The LHC is designed to collide protons with protons in a 26.7~km circumference ring.  Using the roughly 35 pb$^{-1}$ of integrated luminosity delivered at 7 TeV in 2010, both ATLAS~\cite{Aad:2008zzm} and CMS~\cite{Adolphi:2008zzk} experiments measured the top quark pair production cross section.  By 2011, electroweak single top quark production had been observed, and the full complement of top quark measurements and searches for new physics were well underway.  As of the end of $pp$ running in 2011, the LHC has delivered over 5.6~fb$^{-1}$ of integrated luminosity to the ATLAS and CMS experiments, reaching a maximum instantaneous luminosity of $3.6 \times 10^{33}$~cm$^{-2}$s$^{-1}$.  Over the coming years the LHC is expected to gradually increase the instantaneous luminosity to $10^{34}$~cm$^{-2}$s$^{-1}$ or beyond and the center of mass energy up to a maximum of 14~TeV.

Aside from the obvious differences related to the higher center of mass energy and luminosities, there are a few important distinctions between the collider environments at the Tevatron and the LHC.  Perhaps most significantly, because of the higher beam energy, and the lack of valence antiquarks in the LHC beams, gluon initiated top production plays a dominant role at the LHC, and processes involving antiquarks, such as pair production through $q\bar{q}$ annhilation or $s$-channel single top production, have a smaller cross section.  In contrast, at the Tevatron, $q\bar{q}$ annihilation dominates top pair production.  This is particularly relevant for any new physics models where contributions to the top quark sample proceed through $q\bar{q}$ annihilation.  Furthermore, the symmetric nature of the initial state at the LHC means that there is no natural way of defining a ``forward'' or ``backwards'' direction, increasing the challenge of pursuing such measurements as the forward-backward asymmetry $A_{FB}$ in top pair production.  Finally, as the LHC luminosities continue to push to ever increasing levels, the LHC is experiencing never before encountered levels of so-called pile-up from additional $pp$ collisions overlapping with the collision of interest.  These differences in environment can lead to different measurement strategies being produced at the LHC compared to similar analyses performed at the Tevatron.

\subsection{Dominant production mechanism for top quarks}

\begin{figure}
\epsfig{file=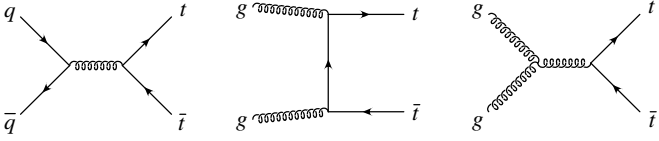,width=.48\textwidth}
\caption{Tree-level LO Feynman diagrams that contribute to $t \bar t$ production.}
\label{fig:ttbar_prod}
\end{figure}

The top quark has color charge, so it can be produced through strong interactions. What makes its cross section so low with respect to generic collider inelastic interactions is its very large mass and the fact that it is produced in pairs via the strong interaction, thus probing high $x$ region of PDFs. Top quark pair production at hadron colliders happens through quark-antiquark annihilation and gluon gluon fusion. The former is dominant at the Tevatron collider where the valence $u$($\bar u$) quark contribution dominates in the proton(antiproton). There are no $\bar u$ valence quarks in the LHC colliding protons, so the corresponding parton density functions (PDF) are very small. On the other hand only a relatively small fraction of the proton's energy is needed to produce top quarks at the 7\,TeV LHC collisions; this energy range is where the gluon PDF dominate. The fraction of top quarks produced through quark\hyp{}antiquark annihilation and gluon\hyp{}gluon fusion is respectively 85\%(15\%) at the Tevatron and 15\%(85\%) at the LHC. The main Feynman diagrams for top quark productions are shown in Fig.~\ref{fig:ttbar_prod}. The total cross section is approximately 7.5\,pb at the 1.96\,TeV $p \bar p$ collisions at the Tevatron, and 160\,pb at the 7\,TeV $pp$ collisions of the LHC~\cite{Moch:2008ai,Aliev:2010zk,Moch:2008qy,Cacciari:2008zb,Kidonakis:2008mu,Langenfeld:2009wd,Kidonakis:2010dk,Ahrens:2010zv,Cacciari:2011hy,Beneke:2009ye}.  For a review summarizing the current status of top production calculations at NLO and with approximate higher-order corrections, see~\cite{Kidonakis:2011ca}.
%

Top quarks can also be produced singly at hadron colliders. This production happens through electroweak diagrams in the $s$- or $t$-channel, or through 
associated production with a $W$ boson. The cross sections at the Tevatron are respectively approximately 1~pb, 2~pb and 0.3~pb~\cite{Harris:2002md,Sullivan:2004ie,Kidonakis:2006bu,Kidonakis:2007wg,Kidonakis:2009mx,Kidonakis:2010tc,Campbell:2009ss}. The rise in the cross section at the LHC is again a function of the number of gluons in the initial state: the $s$-channel production is approximately 5~pb, the $t$-channel is approximately 64~pb, and the $Wt$ production cross sections reaches approximately 16\,pb\,\cite{Kidonakis:2010tc,PhysRevD.83.091503,Kidonakis:2010ux}.

%
\begin{figure}
\epsfig{file=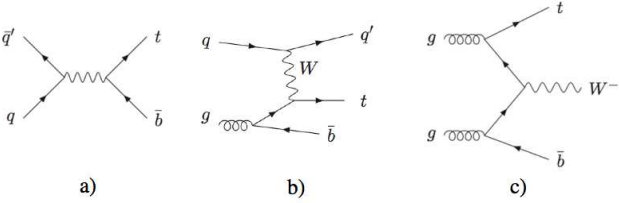,width=.48\textwidth}
\caption{Feynman diagrams for single top quark production. Represented are (a) a LO $s-$channel diagram, (b) a NLO $t$-channel diagram, and (c) a NLO $Wt$ production diagram.}
\label{fig:sttfeyn}
\end{figure}
%

The top quark is the least understood due to the much smaller datasets available when compared to the other quarks. In principle, any sizable deviation in the production rate or in the final state kinematics of top quark events, would be a sign of new physics. There are several theoretically well-motivated models that would predict new physics affecting top quark samples. A fourth generation of heavy quarks is allowed by the SM fit to the existing precision measurement of electroweak observables, and would allow for the right size of CP violation in the universe\,\cite{Holdom:2009rf}.
These exotic quarks would appear in detectors very similarly to events with SM top quark production. Heavy resonances of replicas of the known vector bosons $Z$ and $W$, that appear for example in dynamical electroweak symmetry-breaking schemes\,\cite{topTechnicolor,Hill:1993hs} would affect either or both pair and single top quark production. Supersymmetry theory (SUSY) in combination with the existing collider constraints on SUSY suggest that the supersymmetric partners of the third generation quarks could be the iightest SUSY squarks\cite{Papucci:2011wy}. The production and decay of stop quarks would appear kinematically similar to SM top quark production. 

\subsection{Top quark decay modes}
\begin{figure}
\epsfig{file=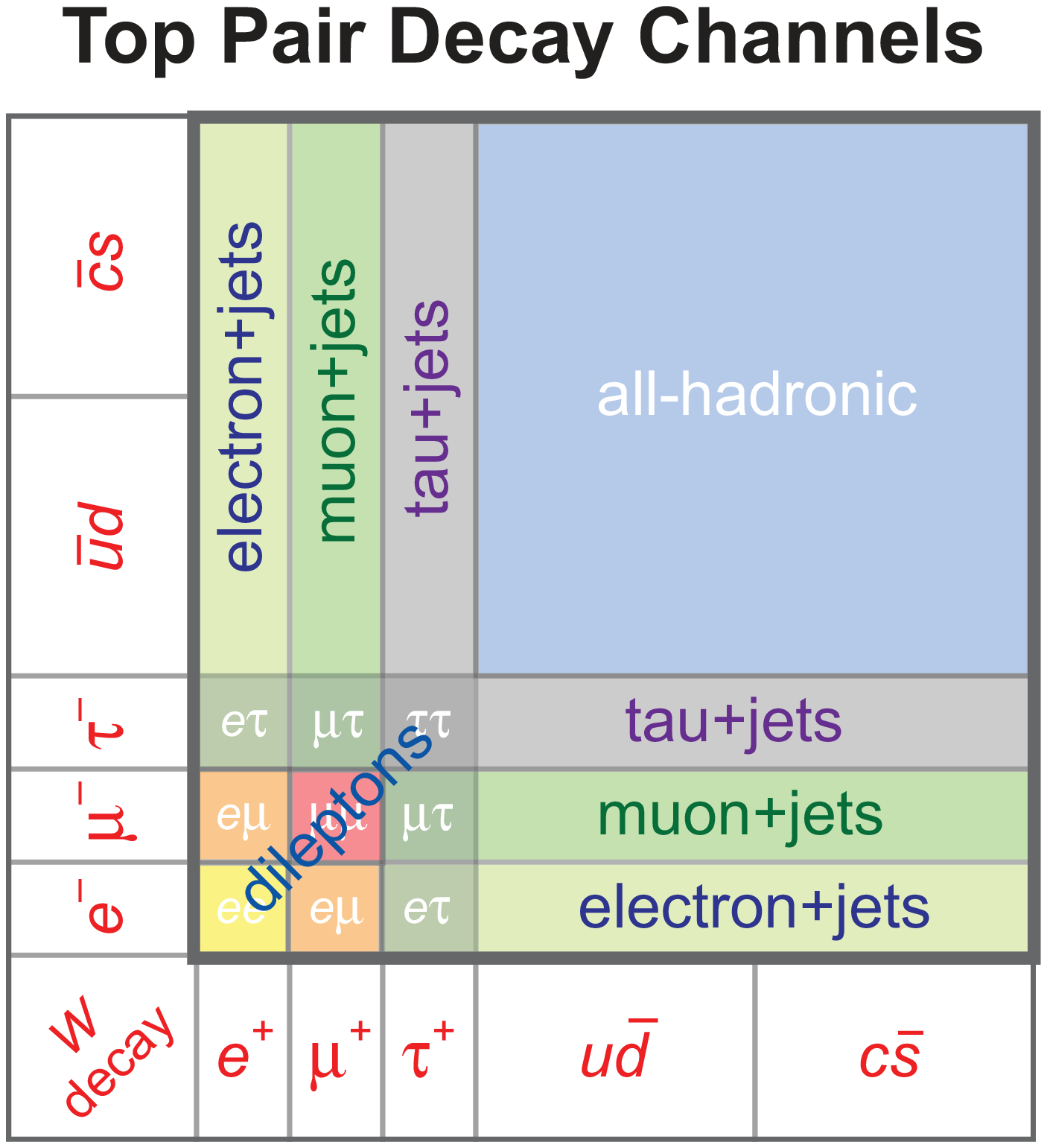,width=.38\textwidth}
\caption{Final states of the $t\bar{t}$ system.}
\label{fig:decay}
\end{figure}

Due to its lifetime being shorter than the hadronization time, the top quark is different from other quarks in that the SM predicts it does not produce resonances. Using precision measurements of CKM parameters and the constraint of its unitarity, the top quark is predicted to decay 99.8\% of the time into a $W$ boson and a $b$ quark. The $W$ boson decays 67.6\% of the time in $u \bar d$ or $c \bar s$\cite{Nakamura:2010zzi} (the conjugate decays implied for the oppositely charged $W$ boson) and the remaining times into a charged lepton $\ell$ and the corresponding neutrino $\nu_\ell$ in the isodoublet. The single top quark decay modes are thus completely specified. Pair production of top quarks leaves a more complex picture: depending on both $W$ boson decays, one is left with a many-quarks final state (``all-hadronic'), a final state composed of four quarks, a charged and a neutral lepton (``lepton+jets") or a final state with two jets, two charged leptons and two neutral leptons (``dilepton''). The relative fractions are 46.2\%, 43.5\% and 10.3\%; a schematic representation is shown in Fig.\,\ref{fig:decay}. 

\section{Analysis Techniques}
\label{sec:AnaTech}

\subsection{Identifying Top Events}
\label{sec:ObjectID}
Top quarks are distinguished from the backgrounds in large part because of the distinctiveness of their event signature, involving some combination of  high energy leptons, jets (including jets originating from $b$ quarks), and a significant amount of missing transverse energy $\etmiss$.  Therefore, the first step of any top quark analysis is identifying the appropriate combination of these signatures.

\subsubsection{Leptons}

In most cases, the term {\it lepton} for a top quark analysis refers specifically to electrons and muons.  Unless explicitly noted, tau leptons are not included, although electrons and muons produced in leptonic tau decays can contribute.  Because they are relatively easy to identify, compared to jets and $\etmiss$, leptons are used in the trigger for many top quark analyses.  The kinematic requirements necessary to produce reasonable trigger rates often determine the lepton acceptance for these analyses.

Dilepton and lepton plus jets analyses at CDF utilize data collected with a trigger involving a single electron or muon with transverse momentum $p_T > 18$ GeV/c within the psuedorapidity\footnote{Psuedorapidity $\eta$ is related to the polar angle with respect to the beam direction $\theta$ as follows: $\eta = - \ln \left[\tan(\theta/2)\right]$.  For a massless particle, $\eta$ is equivalent to rapidity $y$.} range $|\eta| < 2.0$ for electrons and $|\eta| < 1.1$ for muons.  These leptons are then identified offline, requiring $p_T > 20$ GeV, as well as such additional requirements as having a large ratio of electromagnetic to hadronic energy and consistent shower profile for electrons, or a high quality of the track-muon matching for muons.  Additional, non-triggered leptons, such as the second lepton in a dilepton event, or leptons collected on jet- and $\etmiss$-based triggers are allowed to pass looser requirements.  Forward muons ($1.1 < |\eta| < 1.6$) and isolated tracks are sometimes used to increase the muon acceptance.  To suppress leptons from heavy quark decays or hadrons faking leptons, leptons are often required to be isolated, meaning that the ratio between the $E_T$ contained in a cone $\Delta R < 0.4$ around the lepton and the lepton $p_T$ is less than 0.1.

The D0 experiment uses a similar approach:  The D0 single lepton trigger thresholds range from 15 GeV to 80 GeV for electrons and from 10 GeV to 15 GeV for muons.  D0 supplements these single object triggers with a collection of triggers requiring a single lepton with varying $p_T$ thresholds plus a jet.  Offline leptons are selected by requiring $p_T > 20 $ GeV.  Muons are reconstructed in the range $|\eta| < 2.0$ while electrons must have $|\eta| < 1.1$.  Isolated tracks are also used to extend the muon acceptance.

ATLAS and CMS are able to trigger on electrons and muons over a significantly wider $\eta$ range than the Tevatron experiments.  The ATLAS trigger accepts electrons with $E_T > 22$~GeV and $|\eta| < 2.47$, while the muon trigger allows muons with $p_T > 18$~GeV and $|\eta| < 2.5$.  Offline thresholds for electrons and muons are $E_T > 25$ GeV and $p_T > 20$ GeV respectively.  CMS triggers on single electrons ranging from $E_T > 27 $~GeV to 42~GeV, depending on the instantaneous luminosity and whether extra jets are required in the trigger.  For muons, the threshold ranges from $p_T > 17$~GeV/c to 30~GeV/c.  Offline these thresholds range from $E_T > 30 $~GeV to 45~GeV for electrons and $p_T > 20$~GeV/c to 33~GeV/c for muons.  For CMS, the muon acceptance ranges from $|\eta| < 2.1$ to 2.5, while electrons are identified in the range $|\eta| < 2.5$.

\subsubsection{Jets}

Depending on the decay topology, top quark events will have between two and six jets, with up to two of these originating from $b$ quarks.  Identifying energetic jets is key to separating the top signal from the backgrounds.

CDF and D0 use similar jet reconstruction and identification algorithms~\cite{cdfJets,d0Jets}.  Both experiments reconstruct jets with a cone-based algorithm that clusters calorimeter energy deposits into jets.  Jet energies are corrected to account for gain variations across the detector, non-linear energy response, and activity from pile-up.  The jet energies are then calibrated to reproduce the energy of the final state hadrons contained in the jet cone.  Additional corrections to parton-level (in some cases specific to the top quark events) are employed where appropriate (for example, in the top mass measurement).  Both CDF and D0 require jets to have $E_T > 20$ GeV.  D0 jets are reconstructed with a cone size of 0.5 and required to have $|\eta| < 2.5$.  At CDF, jets are reconstructed with a cone size of 0.4 and $|\eta| < 2.0$.

Both ATLAS and CMS reconstruct jets using an anti-$k_T$ algorithm~\cite{anti-kt}.  ATLAS reconstructs jets by clustering together topological energy clusters within the calorimeter above certain energy thresholds using $R = 0.4$.  Energies are corrected from the EM scale to the hadronic scale using $p_T$- and $\eta$-dependent correction factors, and then the absolute jet energy scale is calibrated using test beam data and Monte Carlo simulated collisions.  Jets with $p_T > 25$ GeV and $|\eta| < 2.5$ are considered for top quark analyses.

CMS incorporates tracking information in its jet reconstruction, using one of two algorithms:  The {\it Jet Plus Track} (JPT) algorithm improves jet energy measurements by combining tracking information into jets first reconstructed in the calorimeter alone.  Corrections are made based on the measurements of track momentum both for tracks that project into the area of the jet cone as well as those that project outside.  The second approach taken at CMS is particle-flow jet reconstruction.  The particle flow algorithm attempts to use the full CMS detector to associate all measured energies correctly to one of the following categories: electrons, muons, taus, photons, charged hadrons, and neutral hadrons.  Once the various components are identified, charged and neutral hadrons, plus non-isolated leptons and photons are clustered into jets.  Both jet reconstruction techniques use a size $R = 0.5$.  To be considered for top quark analyses, jets should have $p_T > 30$ GeV and $|\eta| < 2.4$- 2.5.

For $t$-channel single top analyses, it is important to be able to identify and reconstruct jets as far forward as possible, because typically one of the two jets in this signature is in the forward region.  Therefore, single top analyses typically extend the $|\eta|$ of jets used in their analyses.  CDF extends its jet $\eta$ range to $|\eta| < 2.8$, while D0 uses $|\eta| < 3.4$.  ATLAS and CMS extend their jet $\eta$ ranges to 4.5 and 5.0 respectively.

\subsubsection{$b$-Tagging}

Because the top quark nearly always decays into a $W$ boson and a $b$ quark, identifying jets that originate from $b$ quarks is a powerful way to distinguish top production from its backgrounds, as well as to help resolve jet-to-parton matching when reconstructing top quark kinematics.  Algorithms for identifying $b$-quark jets (referred to as $b$-tagging algorithms) typically rely on the long lifetime of the $B$ hadron, either explicitly reconstructing a displaced vertex from the $B$ decay, or by identifying tracks with high impact parameters originating from the $B$ decay.

The CDF experiment primarily uses the {\sc secvtx} $b$-tagging algorithm~\cite{cdfSecVtx} based on reconstructing displaced secondary vertices using the intersection of at least two displaced tracks.  As an alternative, the {\sc JetProb} algorithm~\cite{cdfJetProb} is sometimes used.  In this algorithm, the tracks in a jet are examined and a probability is calculated for all the tracks, based on their impact parameters, to have originated from the primary vertex.  Some CDF analyses also make use of an artificial neural network (ANN) applied after the {\sc secvtx} to increase the purity of the $b$-tagged sample.  The ANN uses input variables related to the tagged secondary vertex, like the decay length, number of tracks, and invariant mass of the tracks associated to the vertex, to discriminate between real $b$-jets and tags of jets that do not originate from a $b$ quark (mistags).  This ANN variable is typically used as an input to other multivariate analysis methods.  The efficiency of the {\sc secvtx} algorithm is approximately 45\% with a mistag rate around 1\%.

D0 uses $b$-tagging algorithm that combines variables from several $b$-tagging approaches using an ANN~\cite{d0NNTagger}.  This tagging algorithm combines the features of three tagging algorithms:  The secondary vertex tagger ({\sc SVT}) is based on reconstructed vertices, while the jet lifetime probability ({\sc jlip}) and counting signed impact parameter ({\sc csip}) taggers are based tracks with high impact parameters.  The {\sc jlip} tagger computes a probability for the jet to originate from the primary vertex, while the {\sc csip} tagger is based on requiring a certain number of high-impact-parameter tracks.   The ANN tagger uses variables related to each of these algorithms, such as the significance of the decay length of the secondary vertex, the invariant mass of the tracks included in the vertex, the {\sc jlip} probability, and the number of {\sc csip} tracks.  By combining information from multiple taggers, the ANN tagger is able to achieve a significantly higher efficiency for a comparable fake rate to the other tagging algorithms.  With this tagger, different operating points may be selected to balance efficiency versus mistag rate.  For example, the D0 single top analysis uses an operating point that yields 47\% $b$-jet efficiency for a mistag rate of 0.5\%~\cite{d0SingleTopPRD}.

The CMS experiment uses three different $b$-tagging algorithms~\cite{cmsBTagger1,cmsBTagger2}.  The simple secondary vertex ({\sc ssv}) tagger reconstructs displaced vertices using two or more charged particle tracks.  The jet probability ({\sc jp}) tagger computes the probability that the tracks in the jet come from the primary vertex, using the impact parameters of the tracks.  A variant of the {\sc jp} algorithm, known as the {\sc jbp} algorithm, gives increased weight to the four tracks with the highest impact parameter.  Finally the track counting ({\sc tc}) algorithm is based on counting the number of tracks with significant impact parameters.  Each of these algorithms has multiple operating points, allowing an optimization of tagging efficiency (which ranges from 36\% to 82\%) and mistag rate (ranging from 0.2\% to 13\%).

ATLAS has implemented two main algorithms for $b$-tagging~\cite{atlasBTagger2,atlasBTagger2}:  The {\sc JetFitter} algorithm is based on reconstructed vertices significantly displaced from the primary vertex.  In contrast, the {\sc ip3d} tagger uses the impact parameters--both longitudinal and transverse--of the tracks in the jet to compute the probability that the jet originates from the primary vertex.  These two algorithms are combined using an ANN to yield an efficiency of 60\% for $b$ jets and approximately a 0.3\% mistag rate.

\subsubsection{Missing Transverse Energy}

Leptonic $W$ bosons decays, associated with a top quark decays, produce energetic neutrinos that cannot be directly detected.  Instead, the presence of these neutrinos have to be inferred by looking at the transverse momentum balance of the visible particles in the detector.  To account for both charged and neutral particles, this momentum balance is usually calculated using energies measured in the calorimeter, weighting the energy in each calorimeter tower by the sine of the polar angle $\sin\theta$.  The vector sum of these weighted calorimeter energies is called the missing transverse energy $\etmiss$.  At CDF and D0, the $\etmiss$ calculated from the raw calorimeter energy deposits, and then corrections are applied based on the calibrated energies for reconstructed objects, like jets, photons, and electrons.  Muons as minimum ionizing particles, are particularly important to correct for because unlike other particles, they do not deposit much of their energy in the calorimeter.  At ATLAS and CMS the $\etmiss$ is calculated starting from calibrated quantities: topological clusters at ATLAS and particle flow candidates at CMS.

\subsection{Signal and Background Modeling}
\label{sec:SigAndBkgModeling}

\subsubsection{Top Pair and Single Top Production}
\label{sec:SigModel}

Top quark production is simulated through a variety of Monte Carlo programs. For top quark pair production, the tree level process is usually described by Leading Order (LO) Monte Carlo simulations  such as {\sc pythia}\,\cite{PYTHIA}, {\sc Alpgen} \,\cite{ALPGEN} and {\sc MadGraph}\,\cite{MADEVENT} respectively by the CDF, D0 and CMS collaborations. The latter two collaborations model $t \bar t$ production through several LO diagrams representing each $t \bar t$ plus zero, one, two or three extra partons; the events are then summed to describe extra radiation at tree level. The ATLAS collaboration uses the Next-to-Leading-Order (NLO) Monte Carlo program {\sc mc@nlo}\,\cite{Frixione:2002ik} to describe $t \bar t$ production. Single top quark production happens through the $s$- and $t$-channel diagrams, or in associated $tW$ production, where the latter is negligible at the Tevatron $p \bar p$ collisions. Its production is modeled by {\sc MadGraph} by the CDF and CMS collaborations and {\sc singletop}\,\cite{Boos:2006af} by the D0 collaboration. ATLAS uses {\sc mc@nlo} to model the single top quark production. Several other Monte Carlo programs have been used either to evaluate possible systematic biases induced by the choice of the default program, or to suit analysis-specific needs.

All tree-level computations are passed to {\sc pythia} for parton shower, hadronization and underlying event, with the exception of {\sc mc@nlo} that is passed to {\sc herwig}\,\cite{Corcella:2000bw} for parton shower/hadronization, and to the subroutine {\sc jimmy}\,\cite{Butterworth:1996zw} for the description of the underlying event. The most common choice for parton distribution functions set (PDF) is the CTEQ one\,\cite{CTEQ5L,Pumplin:2002vw,Nadolsky:2008zw}. Tau decays are simulated through the {\sc tauola}\,\cite{TAUOLA} package. The decayed particles are then passed to a full detector response simulation produced using the {\sc geant}\,\cite{Agostinelli:2002hh} program. Pile-up events are added to the primary collisions through either Monte Carlo simulation or adding real collisions recorded by means of minimum bias triggers.




\subsubsection{Backgrounds}
\label{sec:BkgModel}
In top quark analyses there are two major categories of backgrounds: vector boson ($W$ or $Z$ referred to collectively as $V$) plus jets and multijet QCD.  Both of those background categories are modeled using a different approach.

The predominant technique used to model $V +$ jets production is the matched matrix element plus parton shower (ME+PS) approach.  Exact matrix elements for $V$ plus different numbers of partons are calculated at leading order precision.  Parton-level events generated with these leading-order matrix elements are then fed to a parton shower MC, like {\sc pythia} or {\sc herwig} to account for the effects of parton shower, hadronization, and the decays of unstable particles.  However, there is a double-counting of the phase space that can be populated both by $V+N$ parton events and $V+(N-1)$ parton events with hard radiation from the parton shower.  To remove this double-counting, a matching scheme is needed to veto some events from each sample.  The primary matching scheme used is the MLM matching scheme~\cite{MLM}, but the CKKW scheme~\cite{CKKW,CKKW2} is also sometimes studied.

D0 and CDF use {\sc alpgen} + {\sc pythia} to model $V +$jets production, including matrix elements up to $V + 5$ and 4 partons respectively.  Events including $b$ and $c$ quarks are produced as dedicated samples, including the effects of the heavy quark mass.  ATLAS models $V +$ jets production using {\sc alpgen} + {\sc herwig/jimmy}, again with the events including $b$ and $c$ quarks being generated separately.  At CMS, $V +$ jets processes are modeled with {\sc MadGraph} + {\sc pythia}.  At CMS, heavy flavor quarks are treated as massless and their generation is included in the generation of the rest of the $V+$jets events.

The QCD multijet process is difficult to model with Monte Carlo.  Typically, this process is modeled using one or more side-band regions in the data.  The specific side-band region depends on the top signature being studied.  For example, in lepton + jets analyses, the QCD multijet background is typically modeled using a side-band where the lepton fails one or more selection requirements, like lepton isolation.  For the all hadronic signature, the QCD multijet background is modeled using a sample with relaxed $b$-tagging or kinematic selection requirements.  In the event that a Monte Carlo model for this process is required, typically {\sc alpgen} or {\sc MadGraph} plus {\sc pythia} or {\sc herwig} is used to simulate multijet production.  In some cases, {\sc pythia} or {\sc herwig} dijet production is sufficient.

Additional, electroweak backgrounds as diboson $WW/WZ/ZZ$ production, are usually modeled with {\sc pythia} (at CDF, D0, and CMS) or {\sc herwig} (at ATLAS).  

\subsection{Pair Production}
\label{sec:PairProd}
\subsubsection{Lepton + Jets Final State}

Although commonly referred to as the ``lepton + jets'' final state, this signature generally encompasses only $\mu+$jets and $e+$jets final states.  Final states involving $\tau$ leptons are typically handled separately, as described below.  This signature offers a number of advantages for top quark analyses.  The single energetic charged lepton provides a convenient signal for triggering on these events.  This signature occurs in approximately 30\% of top quark pair production events (neglecting events with taus), offering good compromise between the purity offered by leptonic $W$ decays and the statistics offered by the hadronic $W$ branching fraction.  Because only the $z$-component of the neutrino goes unmeasured, the kinematics of the top quark pair final state can be fully reconstructed by constraining the charged lepton and the neutrino to give the $W$ boson mass, and requiring the same mass for the two reconstructed top quarks.

The primary backgrounds for this signature are irreducible $W+$jets production (with and without additional heavy flavor quarks) and QCD multijet production where one of the several jets fakes a lepton signature.  For typical event selections before $b$-tagging is applied (referred to as the ``pretag'' selection), the signal (S) to background (B) ratio $S:B$ ranges from approximately $1:1$ to $1:5$ depending on the number of jets required and the kinematic selection requirements.  Although the purity in the pretag sample is insufficient to allow a determination of the top quark cross section simply by counting events, it is possible to extract a robust top quark signal by fitting either a single kinematic distribution or multiple kinematic variables combined using a multivariate analysis technique (MVA).  Alternatively, the signal purity can be substantially enhanced by requiring the presence of one or more $b$-tagged jets.  Using $b$-tagging, the $S:B$ can be improved to approximately $2:1$ or more.  The primary challenge for extracting the top signature in the $b$-tagged sample is understanding the rates of $W+$bottom and $W+$charm production.  Theoretical predictions are not sufficiently reliable at this time, so techniques to extract these rates from data have to be used.

\subsubsection{Dilepton Final State}

As with the lepton + jets final state, only electrons and muons are considered as the ``leptons'' of this signature.  Although this signature has the smallest branching fraction (around 5\% after neglecting events with taus), it provides by far the cleanest signature.  The two energetic, isolated charged leptons from the two $W$ bosons decays make this signature easy to trigger on, and the significant amount of $\etmiss$ from two neturinos and the two energetic $b$-jets make this channel easy to separate from the main backgrounds.  Drell-Yan production provides the leading background, and because it is difficult to predict accurately the $\etmiss$ tails, this background is often extracted from data by looking at the $\etmiss$ tails around the $Z$ boson mass peak.  Another challenging background that frequently is extracted from data, is the background for fake leptons.  Events from the $W +$ jets process can be reconstructed in the dilepton final state if one of the jets in the event fakes a charged lepton of opposite sign to the charged lepton from the $W$ decay.  In addition, QCD multijets will contribute if there are two jets providing opposite-sign charged lepton fakes.  Despite these backgrounds, it is common for dilepton event selections to achieve signal to background ratios in excess of $2:1$ with channels like the double-$b$-tagged $e\mu$ sample having virtually no background.  Aside from the lack of statistics, the main challenge of using the dilepton channel comes from the presence of two neutrinos whose four-momenta cannot be measured.
\subsubsection{All-hadronic final state}

This signature has several advantages: the branching ratio BR $\approx 46\%$ is the largest of all decay modes. Also, the jet energy can be measured using only the calorimeter, and the calorimeter coverage is usually broader than the spectrometer coverage for collider detectors, allowing maximal acceptance to the signal. Also, one can in principle reconstruct the full final state kinematics. 

On the other hand, given that it is extremely difficult to identify the charge or flavor of the quark originating the jets, the number of permutation is very large. For the same reason, it is very difficult to discern the top quark from the antitop quark, making several measurements unfeasible in this decay mode. The QCD multijet final state is the most common at a hadron collider, thus isolating the $t \bar t$ signal in this signature requires a detailed understanding of the QCD kinematics and topology. The production of six jets in the final state is poorly understood at theoretical level. It is thus important to utilize the collider data itself to derive a model for the QCD background in this complex final state.

\subsubsection{Events with taus}

Approximately 20\% of top quark pair events appear with third generation leptons in the final state. 
Tau leptons are the most difficult to identify at hadron colliders, due to the multiple ways in which they can appear in the detector. The branching ratio of tau leptons to one or more charged and/or neutral hadron and a tau neutrino is the largest, BR$(\tau \to$hadrons$+\nu_{\tau}) \sim 65\%$. Hadronically decaying taus appear as narrow jets; this signature is easily mimicked by hadronic jets or electrons. The decays of tau into lighter leptons has a lower BR$(\tau \to \ell \nu_{\ell} \nu_{\tau}, \ell=e,\mu) \sim 35\%$ and can hardly be discriminated from electrons or muons produced from $W$ decays. Tau identification algorithms thus address only the hadronic tau decays. 

Due to the large fake rate in tau identification algorithms, the requirement of a tau, large missing transverse energy and jets is not sufficient to produce a sample with reasonable signal purity. The most common choice is thus to identify the additional electron or muon in the dilepton final state, in order to increase purity. In lepton+jets events where the $\ell = \tau$, it is helpful to suppress the dominant QCD background by taking advantage of the different kinematic and topological characteristics of these events, similarly to the what is done in the all-hadronic final state\,\cite{DHare}.

Ultimately, the most effective way to collect top events with taus in the final state has been proven to be by requiring large missing transverse energy, several jets out of which at least one is identified as a $b$-jet, and vetoing the presence of electrons or muons. By exploiting again the peculiar kinematics of the signal events---either by means of a cut based event selection or through a multivariate event selection---it is possible to isolate a kinematic region with large signal purity. 

All of the above choices will leave the remaining top leptonic events as a background to the tauonic signal.


\subsection{Single Top}
\label{sec:SingleTop}
\subsubsection{Lepton + Jets Final State}

Unlike top quark pair production, in single top quark production, there is only one $W$ boson from the top quark decay to provide a charged lepton.  Therefore (except for the associated $tW$ channel which will be discussed momentarily), there is no dileptonic final state, and the lepton + jets final state provides the cleanest event signature.  As in top quark pair production, the lepton + jets signature typically only involves the electron and muon, leaving the tau channel for the $\etmiss + $ jets signature.  For single top quark production, this signature consists of a charged lepton, $\etmiss$ from the neutrino decay, and two to three jets, of which, one or two originate from a $b$ quark.  Because this signature involves fewer jets than top pair production, the backgrounds are substantially more challenging, and extracting the signal without the use of $b$-tagging is not feasible.  Even with $b$-tagging, the typical signal to background ratio in this channel begins at roughly $1:20$ at the Tevatron or $1:7$ at the LHC.  Therefore, extraction of the single top quark production signature typically relies on a combination of $b$-tagging, and multivariate kinematic discriminants.

Because the assocated $tW$-channel involves the decay of an additional $W$ boson produced in association with the top quark, this signature differs slightly from the $t$-channel and $s$-channel production.  The $W$ decay provides either an additional jet or an extra charged lepton and $\etmiss$ compared to the other single top production channels.  This signature is only relevant at the LHC because the production cross section at the Tevatron is negligibly small.
\subsubsection{Events with taus}

The extraction of single top signal from the kinematically similar $W$+jets background has been a major challenge in events with well identified electrons and muons; the isolation of single top events with taus in the final state poses an even greater challenge, as in addition to the W+jets background with real taus, QCD multijet process contribute to the sample whenever a jet originating from a quark or gluon is misidentified as a tau jet. 

Similarly to the analyses of events where top quarks are produced in pairs, leptonically decaying taus are implicitly included in analyses that collect electrons and muons. As of the writing of this document, only CDF and D0 collaboration measured the single top production cross section in events with taus.
Two different strategies have been set forward in both the triggering strategy and the isolation of tau events at the two collaborations.
The CDF experiment analyzes events collected from a trigger path requiring large missing transverse energy from the neutrino, and two energetic jets, while the D0 experiment analyzes events collected from a multitude of triggers typically requiring large energy deposit in the calorimeter due to jet activity. Events with identified electrons or muons are rejected.

The two collaboration implemented also different strategies to address the otherwise dominant QCD background to events with hadronically decaying taus.
The D0 collaboration\cite{Abazov:2009nu}, uses a multivariate tau identification algorithm to suppress the QCD multijet production where a jet mimics a tau. The CDF collaboration\cite{Aaltonen:2010fs} does not attempt to explicitly identify taus, but rather focuses on suppressing the QCD contribution through multivariate techniques, exploiting the different QCD kinematics and topology. In both scenarios, single top decays including electrons and muons will contaminate the tauonic signal.

\section{Experimental Results}
\label{sec:ExpResults}

\subsection{Top Quark Pair and Single Production Cross Section}
\label{sec:XS}

Measurements of the top quark production cross section are good tests
of perturbative QCD.  Deviations in the observed
cross sections from the predictions provided by theory could indicate
the presence of new physics.  Further, precise understanding of the
top-quark pair and single top quark production cross sections enable
searches for new physics in which these processes pose significant
backgrounds.  Hence, measurements of the top-quark production cross
sections are important components of the LHC and Tevatron physics
programs.

At CDF and D0, the $t\overline{t}$ cross section has been measured in
many different channels.  See Table~\ref{tab:tevatron_ttbar_xsecs} for
a summary of a selection of recently published Tevatron results
~\cite{Aaltonen:2010ic,Abazov:2011mi,PhysRevD.82.052002,Abazov:2011cq,PhysRevD.81.052011,PhysRevLett.96.202002}.
Due to channel-dependent background composition and the
$t\overline{t}$ branching ratios to the various accessible final
states, each channel pursued in the measurement of
$\sigma_{t\overline{t}}$ has its own purity and expected yield.  There
is great value in measuring the $t\overline{t}$ production cross
section in multiple channels then, since each attempt must necessarily
approach the measurement in a unique way.  An ultimate combination of
all independent results would have enhanced sensitivity.

\begin{table*}
\begin{center}
  \caption{Summary of published Tevatron results on the $t\overline{t}$
    production cross section.  All measured $\sigma_{t\overline{t}}$
    assume a top-quark mass of 172.5 GeV$/c^2$.}
\begin{tabular}{cccc}
\hline\hline 
Channel          & Experiment & $\mathcal{L}_{\mathrm{int}}(\mathrm{fb}^{-1})$  &  $\sigma_{t\overline{t}} (\mathrm{pb})$ \\ \hline
$\ell$+jets      & CDF        & 4.6  & $7.82 \pm 0.38 \mathrm{(stat)} \pm 0.37 \mathrm{(syst)} \pm 0.15 \mathrm{(Z theory)}$ \\
$(\ell=e,\mu)$   & D0         & 5.3  & $7.78^{+0.77}_{-0.64} \mathrm{(stat + syst + lumi)}$ \\ \hline
dilepton         & CDF        & 5.1  & $7.3 \pm 0.7\mathrm{(stat)} \pm 0.5\mathrm{(syst)} \pm 0.4\mathrm{(lumi)}$ \\ 
$ee,\mu\mu,e\mu$ & D0         & 5.4  & $7.36^{+0.90}_{-0.79} \mathrm{(stat + syst + lumi)}$ \\ \hline
all-hadronic     & CDF        & 2.9  & $7.2 \pm 0.5 \mathrm{(stat)} \pm 1.1 \mathrm{(syst)} \pm 0.4 \mathrm{(lumi)}$ \\ \hline
MET+jets         & CDF        & 2.2  & $7.99 \pm 0.55 \mathrm{(stat)} \pm 0.76 \mathrm{(syst)} \pm 0.46 \mathrm{(lumi)}$  \\ \hline
\end{tabular}
\label{tab:tevatron_ttbar_xsecs}
\end{center}
\end{table*}

Table~\ref{tab:tevatron_ttbar_xsecs} is not intended to be an
exhaustive history of Tevatron Run II $t\overline{t}$ measurements.
The most sensitive $t\overline{t}$ cross section measurements are in
the $\ell$+jets and dilepton channels; hence, the most modern
published results exploiting the largest data samples are listed for
those channels.  Other preliminary high statistics Tevatron results
are available with competitive or superior sensitivity but are not
included here.  Further complicating matters is that some recent Run II
published results make a different assumption for the mass of the top
quark, $m_t$; since the acceptance for $t\overline{t}$ events is
dependent on $m_t$, the measured $\sigma_{t\overline{t}}$ must be
quoted at some assumed $m_t$ value.  Here we chose to only include
results here that make the same assumption ($m_t=172.5$ GeV$/c^2$) as
the main $\ell$+jets and dilepton analyses to facilitate comparison.
Lastly some results having large uncertainties relative to the results
from the $\ell$+jets and dilepton channels, these were not included
either.

Several qualitative features of these results can be identified:  First, the
NLO prediction for the $t\overline{t}$ production cross section at the
Tevatron has been recently calculated by Moch and Uwer
~\cite{Moch:2008ai}: $\sigma^{\mathrm{NLO}}_{t\overline{t}}=7.5 \pm 0.7$
pb.  The results listed in Table~\ref{tab:tevatron_ttbar_xsecs} are
all completely consistent with this prediction from theory.  Each
measurement listed is consistent with the NLO prediction within 
1$\sigma$ uncertainties.

The most precise measurement of $\sigma_{t\overline{t}}$
~\cite{Aaltonen:2010ic} is extracted from a measurement of the ratio
$R=\sigma_{t\overline{t}}/\sigma_{Z}$.  In the measurement of $R$,
several sources of systematic uncertainty cancel; one can then exploit
the superior precision of the theoretical $Z$ production cross section
to achieve a measured $\sigma_{t\overline{t}}$ with significantly
reduced systematic uncertainty over conventional methods.  Two such
measurements were executed at CDF; the most precise measurement
discarded $b$-tagging information, due to the systematic uncertainties
one incurs when using tagging variables, and exploited a neural
network for final event classification, examining the kinematic
variables of the events. This superior Tevatron
$\sigma_{t\overline{t}}$ result is now statistics limited; this
measurement will therefore benefit from an updated result exploiting
the full Tevatron Run II statistics.  The remaining systematic
uncertainty is comparable to the statistical uncertainty, so it is
important to keep the systematic uncertainty in mind for future
analyses; dominant remaining sources of systematic uncertainty include
the uncertainty on the scale for jet energy measurements and
uncertainty in the modeling of signal $t\overline{t}$ and background
$W$+jets events.

The best Tevatron measurement of $\sigma_{t\overline{t}}$ has a relative 
total uncertainty of $\sim 7$\% .  Note that the NLO theoretical prediction
has a relative uncertainty of  $\sim 9$\% .  Hence, the precision of the
measurement of the $t\overline{t}$ is now exceeding that of the theoretical
prediction; to identify new physics through an observed 
discrepancy in the measured $\sigma_{t\overline{t}}$ with respect to theory,
then new, more precise theoretical predictions will be necessary.  While
approximate higher order calculations exist, no complete NNLO is currently
available.  Hence, many in the top-quark physics community are eager for 
the measurement of the differential production cross sections of 
top-quark pairs in the high-statistics samples at the LHC experiments
to look for possible indications of new physics.

As discussed above, the main production mechanism for top-quark pairs
at the Tevatron is quark-antiquark annihilation, whereas at the LHC
top-quark pairs come mostly through gluon-gluon fusion.  Hence, in
addition to simply probing top-quark production at a higher
center-of-mass collision energy, measurements of the $t\overline{t}$
production cross section at the LHC also test our understanding of the
top-quark pair production mechanism in a fundamentally new regime.
Additionally, early LHC measurements
~\cite{Khachatryan:2010ez,Aad:2010ey} of the $t\overline{t}$
production cross section were used to demonstrate the health of the
LHC and the general purpose experiments CMS and ATLAS.

Table~\ref{tab:lhc_ttbar_xsecs} contains a summary of the published
measurements of $\sigma_{t\overline{t}}$ from CMS and ATLAS using the
full 2010 data sample~\cite{Chatrchyan:2011ew,Chatrchyan:2011yy,Chatrchyan:2011nb,Aad:2011yb}.  
New analyses exploiting the full statistics of the 2011 LHC
run remain preliminary~\cite{ATLAS-CONF-2011-140,ATLAS-CONF-2011-121,ATLAS-CONF-2011-100,CMS-PAS-TOP-11-024,CMS-PAS-TOP-11-005,CMS-PAS-TOP-11-003,CMS-PAS-TOP-11-007,CMS-PAS-TOP-11-006}.  

\begin{table*}
\begin{center}
  \caption{Summary of published LHC results on the $t\overline{t}$
    production cross section.  All measured $\sigma_{t\overline{t}}$
    assume a top-quark mass of 172.5 GeV$/c^2$.}
\begin{tabular}{cccc}
\hline\hline 
Channel          & Experiment & $\mathcal{L}_{\mathrm{int}}(\mathrm{fb}^{-1})$  &  $\sigma_{t\overline{t}} (\mathrm{pb})$ \\ \hline
$\ell$+jets - kinematics only & CMS        & 0.036  & $173^{+39}_{32} \mathrm{(stat. + syst.)} \pm 7 \mathrm{(lumi)}$ \\
$\ell$+jets - with $b$-tagging& CMS        & 0.036  & $150 \pm 9 \mathrm{(stat)} \pm 17 \mathrm{(syst)} \pm 6 \mathrm{(lumi)}$ \\ \hline
dilepton         & CMS        & 0.036  & $168 \pm 18 \mathrm{(stat)} \pm 14 \mathrm{(syst)} \pm 7 \mathrm{(lumi)}$ \\
                 & ATLAS      & 0.035  & $171 \pm 20 \mathrm{(stat)} \pm 14 \mathrm{(syst)} ^{+8}_{-6} \mathrm{(lumi)}$ \\ \hline
\end{tabular}
\label{tab:lhc_ttbar_xsecs}
\end{center}
\end{table*}

Similar qualitative observations can be made regarding these early LHC
as were made for the summary of Tevatron Run II results.  The
prediction from NLO theory indicates that
$\sigma^{\mathrm{NLO}}_{t\overline{t}}=157 \pm 24$ pb~\cite{Moch:2008ai,Aliev:2010zk,Moch:2008qy,Langenfeld:2009wd};
each of the results in Table~\ref{tab:lhc_ttbar_xsecs} is consistent
with this NLO prediction within 1$\sigma$ total uncertainty.  The best
current measurement at the LHC comes from the $\ell$+jets channel at
CMS in an analysis that exploited $b$-tagging in event classification.
This technique extracted the $t\overline{t}$ content of the selected
sample through a profile likelihood fit to the number of total
reconstructed jets, the number of $b$-tagged jets, and the secondary
vertex mass distribution in the data.  The main systematic
uncertainties are taken into account when maximizing the profile
likelihood, hence the systematic uncertainty on the measured
$t\overline{t}$ cross section is reduced with respect to more
conventional techniques.  Under these conditions, this result
achieves a sensitivity to $\sigma_{t\overline{t}}$ comparable to that 
of NLO theory.  

All these 2010 LHC results will soon be eclipsed in terms of
sensitivity by the results from the full 2011 LHC data sample.  At the
time of preparation of this Review, all of the 2011 remain
preliminary.  With the high statistics 2011 data samples, reduction of
systematic uncertainty will be the top priority in all measurements of
$\sigma_{t\overline{t}}$.

After discovery of the top quark through the strong interaction and
subsequent measurement of the top quark's mass, electroweak production
of single top quarks became the next major goal in top-quark physics.
Single top quark production is not just a curiosity; the single top
cross section is sensitive to the CKM matrix element $|V{tb}|$,
providing the opportunity for first direct measurement of this
parameter within the SM.  

At the Tevatron, as discussed above, single top quarks come through
two production mechanisms, s- and t-channel production.  In early
searches, the s- and t-channel mechanisms were searched for together,
looking for evidence of their combined contribution.  The predicted
$(s+t)$ channel production cross section at the Tevatron is 3.0-3.5
pb~\cite{Harris:2002md,Sullivan:2004ie,Kidonakis:2006bu,Kidonakis:2007wg,Kidonakis:2009mx,Kidonakis:2010tc,Campbell:2009ss}; although this is roughly half 
the inclusive $t\overline{t}$ production cross section, the task of
extracting the single top signal is made significantly more
challenging by the presence of significantly more backgrounds in the jet
multiplicity bins in which the signal resides, compared to the relevant jet bins for
top-quark pair production.

First observation of single top quark production was achieved in 2009
by both CDF~\cite{cdfSingleTopObs,Aaltonen:2010jr} and D0~\cite{d0SingleTopObs}.  These
results were combined~\cite{Group:2009qk} yielding a measured CDF+D0
single top cross section of

\begin{equation}
\sigma_t=2.76^{+0.58}_{-0.47} \mathrm{(stat + syst)} \mbox{ pb}, 
\end{equation}

\noindent assuming $m_t=170$ GeV$/c^2$, completely consistent with
the prediction from the SM.   This measured cross section corresponds to 
a measurement of 

\begin{equation}
|V_{tb}|=0.88 \pm 0.07 \mathrm{(stat + syst)},
\end{equation}

\noindent corresponding to a 95\% C.L. lower limit of $|V_{tb}|>0.77$.  These
results on $|V_{tb}|$ are also consistent with the expectation from
the SM ($|V^{\mathrm{SM}}_{tb}| \sim 1.0$).  

D0 has also measured single top production using explicitly the $\tau+$jets signature~\cite{Abazov:2009nu}, extracted for the first time the $t$-channel cross section separately with a model independent technique~\cite{Abazov:2011rz}, and updated the single top cross section measurement to 5.4~fb$^{-1}$~\cite{Abazov:2011pt}.  Each of these results is consistent with SM expectations for electroweak single top production. A review of Tevatron single top quark results can be found in\,\cite{Heinson:2011jn}.

At the LHC, the early focus has been on establishing t-channel single top 
production directly.  The CMS experiment performed the first measurement
of t-channel single top production~\cite{Chatrchyan:2011vp} at the LHC
in the 36 pb$^{-1}$ 2010 data sample obtaining a $t$-channel cross section---summing $t$ and $\bar{t}$ contributions---of $\sigma = 83.6 \pm 29.8 \mbox{ (stat.+syst.)} \pm 3.3 \mbox{ (lumi.)}$~pb .  Preliminary results from CMS on a search for the $tW$ channel production~\cite{CMS-PAS-TOP-11-022} and ATLAS  on a search for the $s$-channel production~\cite{ATLAS-CONF-2011-118} and a measurement of $t$-channel production~\cite{ATLAS-CONF-2011-101} yield results that are so far consistent with the standard model.  With the increased datasets available in 2011 and 2012 the LHC experiments will continue to expand the range of single top quark measurements.

%
%
%

\section{Measurement of the top quark mass}
\label{sec:Mass}

The top quark mass is a free parameter in the standard model of particle physics and must thus be experimentally determined. The top quark mass gives large contribution to electroweak radiative corrections, and can be used together with other electroweak observables to infer the Higgs boson mass in both SM and non-SM scenarios\,\cite{Baak:2011ze}. The improvement in the precision on the top quark mass measurement in the recent years has narrowed significantly the mass range for the existence of the SM Higgs boson. Also, the precise measurement of this parameter is of crucial importance as it determines many of the other properties of the top quark: as an example, the dependence of the theoretical computation of $\sigma_{t \bar t}$ from the top quark mass is $\approx 3\%/$(GeV/$c^2$).

Measuring the top quark mass requires a large statistics top quark sample, sophisticated analysis tools, and excellent understanding of the detector response and of the physics of $t \bar t$ events\,\cite{Galtieri:2011yd}.
The current most precise estimation\,\cite{TevatronElectroweakWorkingGroup:2011wr} of the top quark mass comes from the combination of the CDF and D0 results in the lepton+jets\,\cite{Aaltonen:2010yz,Abazov:2011ck} and dilepton channels\,\cite{Aaltonen:2011dr,Abazov:2011fc}, from the CDF measurements in the all hadronic\,\cite{Allhad} and MET+jets\,\cite{METjets} channels, and a measurement that is largely independent of the jet energy scale\,\cite{Aaltonen:2009hd}. The average of the above measurements using the 2001-2009 Tevatron dataset, and of the earlier results using the 1992-1996 Tevatron data, gives $M_{top} = 173.2 \pm 0.6\mbox{ (stat)} \pm 0.8 \mbox{ (syst)}$\,GeV/$c^2$, corresponding to a 0.56\% uncertainty. The largest systematic source comes from the uncertainty on the signal modeling. As the systematics include a the JES term that scales with luminosity, a precision below 0.5\% is achievable, once the Tevatron data accumulated in the 2010-2011 running is incorporated.

\begin{figure}[htbp]
\includegraphics[width=0.40\textwidth]{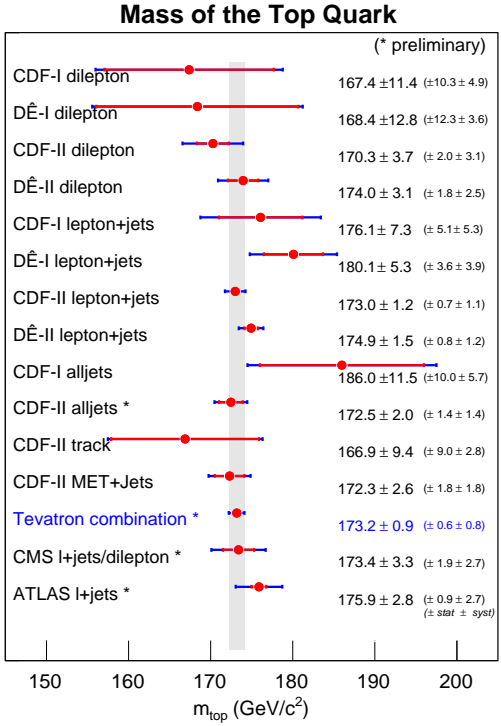} 
\caption{Summary of the Tevatron and LHC measurements of the top quark mass.}
\label{fig:Mtop} 
\end{figure}

The CMS collaboration measured the top quark mass using the 2010 LHC dataset in the lepton+jets\,\cite{CMSljets} and dilepton\,\cite{Chatrchyan:2011nb} final states, finding good agreement with the Tevatron result. Using 0.7\,fb$^{-1}$ of data in the lepton + jets channel, ATLAS measures $M_{top} = 175.9 \pm 0.9 \mbox{  (stat)} \pm 2.7 \mbox{ (syst)}$\,GeV/$c^2$\,\cite{ATLASljetsnew}. The precision is currently limited by the limited understanding of the jet energy scale, the initial and final state radiation, and tree level modeling uncertainties. The Tevatron and LHC measurements and their overall agreement can be seen in Fig.\,\ref{fig:Mtop}. 

All the mentioned above are calibrated to Monte Carlo simulations. Thus the mass measured is effectively the mass definition contained in the LO Monte Carlo used; theorists agree that the Monte Carlo mass should be very close to the top quark pole mass. Beyond LO QCD, the mass of the top quark is a convention-dependent parameter, the other dominant convention being the MS scheme. To probe further into this ambiguity, the D0 and ATLAS collaborations compared the measured inclusive $t \bar t$ production cross section with fully inclusive calculations at higher-order QCD that involve an unambiguous definition of $M_{top}$ and compares the results to MC\,\cite{Abazov:2011pta,ATLASxsec}. Both measurements favor the pole mass hypothesis over the $\bar MS$ hypothesis.

\begin{figure}
 \begin{centering} 
\includegraphics[width=.34\textwidth]{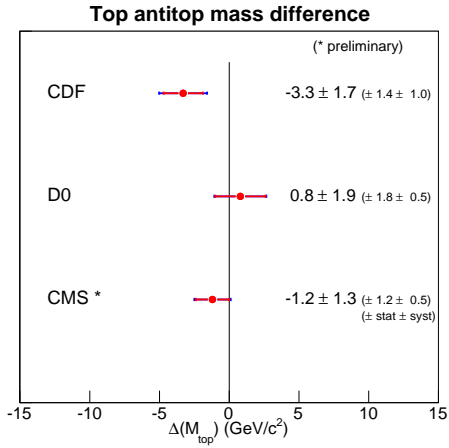}
\caption{Summary of the Tevatron and LHC results on the measurement of the difference among the top and antitop quark mass.}
\label{fig:DeltaMtop}
\end{centering} 
\end{figure}

Due to the shortness of the top quark lifetime, the top quark is the only quark that can be studies before hadronization occurs; this fact allows a unique opportunity to measure directly the mass difference between a quark and its antiquark as a test of the CPT symmetry conservation\,\cite{Cembranos:2006hj}. The measurement of the difference between the top quark and antitop quark mass relies on the same techniques that have been developed to measure the top quark mass. The advantage here is that almost all systematics affecting the $M_{top}$ measurement cancel out in the $\Delta(M_{top})$ measurement as they affect the measurement of $M_{top}$ and $M_{antitop}$ in a highly correlated manner.
The CDF\,\cite{Aaltonen:2011wr}, D0\,\cite{Abazov:2011ch} and CMS\,\cite{CMSmassdiff} collaborations measured this difference to be in agreement with the standard model preciction of no difference, to a precision up to $\frac{\Delta M_{top}}{M_{top}} = 0.7\%$. A summary of the existing measurements is  shown in Fig.\,\ref{fig:DeltaMtop}.

\subsection{Measurement of the top quark width}
\label{sec:Width}
In the SM, the total decay width of the top quark $\Gamma_t$ is dominated by the partial decay width $\Gamma(t \to Wb)$.
The top quark partial width $\Gamma(t \to Wb)$ has been computed to be  $\Gamma(t \to Wb) = \frac{G_F M_{top}^3}{8 \pi \sqrt{2}} |V_{tb}|^2$. 
The top quark width $\Gamma_t$ ranges from 1.26 to 1.4 depending on the top quark mass used (170-175). A measurement of this quantity would test pQCD and EWK calculations, while deviations from this value could be induced by decays of top quarks to non-SM particles such as scalar top partners, charged Higgs, or FCNC decays. The decay width of an unstable particle can be measured in principle from its mass spectrum. The width of the reconstructed top quark mass distribution is dominated by the jet energy resolution, and is tipically of the order of tens of GeV at current experiments. While $\Gamma_t$ is far smaller than the experimental resolution on $m_t^{reco}$, the shape of the mass spectrum retains some dependence on the top quark width. The CDF collaboration utilized the top quark mass spectrum to measure $\Gamma_t$. The {\em in situ} calibration of the jet energy scale using the hadronically decaying $W$ boson in performed to suppress the otherwise dominant jet energy scale uncertainty\,\cite{Aaltonen:2010ea}. Using 4.3\,fb$^{-1}$ of $p \bar p$ collisions, CDF reports the 68\%CL interval to be $0.3 < \Gamma_t < 4.4$\,GeV. A more precise measurement of the total width can be obtained from the equation $\Gamma_t = \Gamma(t \rightarrow W b) / BR(t \rightarrow W b)$, where the partial width $\Gamma(t \rightarrow W b)$ is measured through the total production cross section of single top events, and $BR(t \to Wb)$ is measured in top quark pair events\,\cite{Yuan:1995sh}. This technique assumes that the $tWb$ coupling leading to single top quark production is identical to the coupling leading to top quark decay. The D0 collaboration used as input the measurement of $BR(t \to Wb)$ that utilized $\sim 1\,$fb$^{-1}$ of $p \bar p$ collision data\,\cite{Abazov:2008yn}, and the measurement of the $t-$channel single top quark cross section that uses 2.3\,fb$^{-1}$ of data\,\cite{Abazov:2009pa} to measure $\Gamma_t = 1.99^{+0.69}_{-0.55}$\,\cite{Abazov:2010tm}.
It is interesting to note that the D0 collaboration recently released a newer and more precise result for the measurement of $BR(t \to Wb)/BR(t \to Wq)$ that utilizes five times more data, observing a value of $0.90\pm0.04$\,\cite{Abazov:2011zk}. This measurement corresponds to approximately a 2.5\,$\sigma$ deviation from the SM prediction: it will thus be interesting to see the same measurement performed by the CDF, ATLAS and CMS collaborations. 



\subsection{$W$-Helicity}
\label{sec:WHelicity}
As the only quark that decays before hadronization, the top quark offers a unique opportunity to study directly the electroweak interaction.  According to the SM, the V-A structure of the $Wtb$ vertex leads to the prediction that the $W$-boson helicity in top decays will be 69.8\% longitudinal, 30.1\% left-handed, and 0.041\% right-handed~\cite{AguilarSaavedra:2006fy,Cao:2003yk,delAguila:2002nf,Kane:1991bg}.  Any deviation from these expected values would be a sign of new physics modifying the $Wtb$ vertex.

Currently the most precise constraints on the $W$ boson helicity fraction in top decays comes from the combination of CDF and D0 measurements from the Tevatron~\cite{cdfD0WHelComb}.  This combination uses three separate measurements:
\begin{description}
\item[CDF lepton+jets:]  This measurement uses 2.7 fb$^{-1}$ of integrated luminosity~\cite{cdfWHelLJ}.  The purity of the selected data sample is enhanced by requiring a $b$-tagged jet as part of the event selection.  The $W$ boson helicity fractions are extracted from the data with the help of a Matrix Element technique that attempts to make use of as much kinematic information from the event as possible, integrating over poorly known quantities.  Leading-order matrix elements for both the top pair production signal and the main backgrounds are included.
\item[CDF dilepton:]  This measurement uses 5.1 fb$^{-1}$ of integrated luminosity~\cite{cdfWHelDIL}.  Selected events are divided into ``tagged'' and ``untagged'' subsamples based on the presence of a $b$-tagged jet, and all events are used.  The $W$ boson helicity fractions are extracted using the cosine $W$ boson helicity angle $\cos\theta^*$, which is defined by the angle between the down type quark from the $W$ boson decay and the opposite of the top quark direction in the $W$ boson rest frame.  To reconstruct the undetected neutrino kinematics, both the $W$ boson and top quark mass constraints are utilized, and experimental resolution on the detected leptons, jets, and missing transverse energy are accounted for.  Ambiguities in the solution coming from jet-to-lepton pairs and kinematics are resolved by taking the most probable solution (considering detector resolution) that gives the smallest effect invariant mass for the $t\bar{t}$ system.
\item[D0 lepton+jets and dilepton combined:]  This result uses data corresponding to 5.4 fb$^{-1}$ of integrated luminosity from both the lepton + jets and dilepton channels~\cite{d0WHelLJDIL}.  For the lepton + jets data, events are selected with at least four jets, and two of them are required to be $b$-tagged.  The signal purity of both the lepton + jets and the dilepton channels is enhanced using a multivariate likelihood variable to select a subsample enhanced in top quark content.  As with the previous measurement, the $W$ boson helicity fractions are extracted from the $\cos\theta^*$ distribution.  For the lepton + jets channel, the undetected neutrino is reconstructed using a kinematic fit incorporating both $W$ boson mass and top quark mass constraints.  In the dilepton channel, the reconstruction of the two undetected neutrinos is accomplished using a technique similar to the CDF dilepton measurement.
\end{description}

The individual results are combined using the best linear unbiased estimator (BLUE) technique~\cite{BLUE1,BLUE2} which is capable of accounting for correlations in the systematic uncertainties.  The results are calculated two ways: (1) the model independent (or 2D) approach where no assumptions are made about the individual helicity fractions beyond the requirement that they sum to 1, and (2) the model dependent (or 1D) approach where either the longitudinal ($f_0$) or right-handed ($f_+$) is assumed to have the SM expected value, and the value of the other quantity is extracted (again, assuming the sum of all fractions is 1).  Table~\ref{tab:tevWHel} summarizes the results obtained.

\begin{table}[htdp]
\caption{Summary of the Tevatron $W$ helicity combination.  Both the longitudinal fraction $f_0$ and right-handed fraction $f_+$ are reported.}
\begin{center}
\begin{tabular}{ccc}
\hline\hline
Method                                & $f_0$                      & $f_+$ \\
Model independent (2D) & $0.732 \pm 0.081$ & $-0.039 \pm 0.045$ \\
Model dependent (1D)    & $0.685 \pm 0.057$ & $-0.013 \pm 0.035$ \\
\hline\hline

\end{tabular}
\end{center}
\label{tab:tevWHel}
\end{table}%

ATLAS has also measured the $W$ helicity using both lepton+jets and dilepton events in data corresponding to 0.7 fb$^{-1}$ of integrated luminosity~\cite{atlasWHelLJDIL}.  Lepton + jets events were required to have at least one $b$-tagged jet, while no $b$-tagging requirement was applied to the dilepton channel.  For lepton + jets events, the neutrino kinematics were reconstructed using a constrained kinematic fit, while the        neutrinos in dilepton events were reconstructed solving a set of constraint equations, choosing the solution that gives the smallest product of transverse neutrino momenta.  The $W$ helicity fractions are extracted from the $\cos\theta^*$ distribution.  The results obtained are listed in Table~\ref{tab:atlasWHel}.

\begin{table}[htdp]
\caption{Summary of the ATLAS $W$ helicity measurements.  Results are quoted for the longitudinal fraction $f_0$, the left-handed faction $f_-$ and right-handed fraction $f_+$ are reported.}
\begin{center}
\begin{tabular}{cccc}
\hline\hline
Method                                & $f_0$                        &       $f_-$                   & $f_+$ \\
Single Lepton & $0.57 \pm 0.11$     &  $0.35 \pm 0.06$    & $0.09 \pm 0.09$ \\   
Combined       & $0.75 \pm 0.08$     & $0.25 \pm 0.08$     & (Assumed zero) \\
\hline\hline

\end{tabular}
\end{center}
\label{tab:atlasWHel}
\end{table}%

In addition to extracting the $W$ helicity fractions, the ATLAS data are used to extract limits on anomolous couplings at the $Wtb$ vertex.

\subsection{Spin Correlations}
\label{sec:SpinCorr}

Although it is not possible to measure directly the spins of
pair-produced top quarks at hadron colliders, the individual spins of
the top and anti-top quarks are predicted to be significantly
correlated in the SM~\cite{Barger:1988jj}.  Additionally, within the
SM, the top quark decays via the electroweak interaction $t
\rightarrow Wb$ long before hadronizing through the strong
interaction.  Hence, the angular distributions of the final state
particles in $t\overline{t}$ production will depend on the spin
orientations of the parent $t$ and $\overline{t}$
quarks~\cite{Bernreuther:2004jv}.  One can use these angular
distributions as a probe of the spin correlation of top-quark pairs
and check for consistency with what is expected in the SM.
Significant de-correlation would indicate that the spins of the top
quarks had flipped before decay -- or that the spin orientation
information was not propagated as expected to the top quark decay
products, such as would occur in the non-SM decay $t \rightarrow
H^+b$.

The $t\overline{t}$ spin correlation parameter $C$ is defined by
\[\frac{d^2\sigma}{d\cos \theta_1 d\cos \theta_2}=\frac{\sigma(1-C\cos \theta_1\cos
\theta_c)}{4},\]
where $\sigma$ denotes the $t\overline{t}$ cross section
and $\theta_i$ denotes the angles between the chosen spin-quantization
axis and the direction of flight of the down-type fermion in the $W$
decays, measured in the appropriate rest frame of the $t$ or
$\overline{t}$.  The value $C=+1 (-1)$ corresponds to fully correlated
(anti-correlated) spins, and $C=0$ corresponds to no spin correlation.
Within the SM, for dilepton channel $t\overline{t}$ production at the
Tevatron, NLO QCD theory predicts
$C=0.78^{+0.03}_{-0.04}$~\cite{Bernreuther:2004jv}; the prediction for $C$
in other $t\overline{t}$ channels will require a small correction
due to because of the different spin analyzing power between leptonic
and hadronic $W$ decays~\cite{Brandenburg:2002xr}.

Several spin correlation parameter $C$ measurements have been
attempted at the Tevatron (see for example
\cite{Abazov:2011qu,Aaltonen:2010nz} for recent published results).
None of these measurements, nor any of the measurements from earlier
in the Run II Tevatron physics program, had the necessary sensitivity
to measure $C$ with sufficient precision so as to be able to
discriminate between the no-correlation and SM-level correlation
hypotheses.  Recently, however, the D0 Collaboration
performed a measurement in a data sample corresponding to 5.4~fb$^{-1}$
of integrated luminosity that had sufficient sensitivity to eliminate
the no-correlation hypothesis~\cite{Abazov:2011gi}.  

The technique relies on leading-order (LO) matrix elements (ME) to
measure the ratio $f$ of events with correlated $t$ and $\overline{t}$
spins to the total number of $t\overline{t}$ events.  It follows then
that $f=1 (0)$ corresponds to complete correlation (zero
correlation). The event-by-event $t\overline{t}$ signal probability
$P_{\mathrm{sgn}}$ can be calculated from the LO MEs for both the
correlated and uncorrelated spin hypotheses.  One can write
$P_{\mathrm{sgn}}$ as

\begin{align*}
P_{\mathrm{sgn}}(x;H)= & \frac{1}{\sigma_{\mathrm{obs}}}\int F_{\mathrm{PDF}}(q_1)F_{\mathrm{PDF}}(q_2)\mathrm{d}q_1\mathrm{d}q_2 \\
 & \times \frac{(2\pi)^4|\mathcal{M}(y,H)|^2}{q_1q_2s}W(x,y)\mathrm{d}\Phi_6,
\end{align*}

\noindent where $\sigma_{\mathrm{obs}}$ is the LO $q\overline{q}$
annihilation $t\overline{t}$ production cross section corrected to
include acceptance and selection efficiency effects, $q_i$ is the
fraction of the incoming proton or antiproton that parton $i$
possesses, $F_{\mathrm{PDF}}(q_i)$ is the parton distribution function
of the $i$-th parton, $\mathrm{d}\Phi_6$ is the infinitesimal volume
in the 6-body kinematic phase space and $s$ is the square of the
center-of-mass energy in the colliding beams.  $W(x,y)$ is a mapping
function that contains the probability that a partonic system of $n$
leptons and jets described by the partonic final state $y$ could
manifest itself in the measured event as the four-momenta collection
$x=(\vec{p_1},\vec{p_2},...\vec{p_n})$.  Two different hypotheses
are considered, $H=$correlated or uncorrelated; $\mathcal{M}(y,H)$
are the LO MEs calculated for the two different hypotheses.

The event-by-event discriminant $R$ is then defined as

\begin{equation*}
R(x) = \frac{P_{\mathrm{sgn}}(x;correlated)}{P_{\mathrm{sgn}}(x;correlated)+P_{\mathrm{sgn}}(x;uncorrelated)}.
\end{equation*}

\noindent Templates from the distribution $R$ are constructed from MC
for $t\overline{t}$ with spin correlations, without spin correlations
and for the largest backgrounds.  A multi-component binned
maximum-likelihood fit is then performed in the data; from the resulting
yields one can calculate $f$.  

The measurement is performed independently in the lepton+jets and
dilepton channels and the individual results are combined.  The resulting
$f$ is measured as

\begin{equation}
f = 0.85 \pm 0.29 (\mathrm{stat}+\mathrm{syst}).
\end{equation}

\noindent Recall that $f=0$ corresponds to zero correlation among the
$t$ and $\overline{t}$ quarks.  In addition to measuring $f$, one can
alternatively set a lower limit on its range; the values
$f<0.344(0.052)$ are accordingly excluded at 95\% (99.7\%) CL.
Hence, through this measurement, the zero spin correlation hypothesis
in $t\overline{t}$ production is ruled out for the first time.  The
result still suffers from large uncertainties; as it is statistically
limited, updated spin correlation results from both Tevatron
experiments exploiting the full Run II data sample would be very
welcome.

It should be noted that the strength of the spin correlation is
predicted to be different at NLO for $q\overline{q}$ annihilation and
$gg$ fusion production of top-quark pairs~\cite{Mahlon:1997uc}.
Hence, given the dominance of $gg$-fusion-induced $t\overline{t}$ at
the LHC, measurements at CMS and ATLAS provide an important complement
to the suite of Tevatron measurements, yielding new insight top-quark pair production.  Top-quark spin
correlation studies have been undertaken in the early data samples at
the LHC~\cite{Aad:temp}.  The techniques employed
for measuring the spin correlation are qualitatively similar to those
employed in the Tevatron analyses.  These early LHC results have not
yet been published at the time of the preparation of this Review;
however, the preliminary results from the LHC indicate that the
observed level of spin correlation between top-quark pairs at the LHC
is not inconsistent with SM expectations.  

%
%
%

\subsection{Resonance Searches}
\label{sec:ResSearch}
A common new physics signature involving top quarks is a heavy resonance that decays to 
top quark pairs.  One way to detect such a resonance is by it's effect on the $t\bar{t}$ invariant 
mass distribution, either directly through a resonant bump, or through the distortion caused by 
interference between the new physics and the Standard Model.

One search strategy employed at both the Tevatron and LHC experiments is to seek evidence of a bump in the reconstructed $t\bar{t}$ invariant mass distribution, indicating the presence of a narrow width resonance.  The reference model generally used for such a search is 
a leptophobic $Z^{\prime}$ boson that decays preferentially to top quark pairs with a width 
$\Gamma_{Z^{\prime}} = 0.012 M_{Z^{\prime}}$~\cite{narrowZPrime}.  However, several other 
models are also tested, including ones involving wider leptophobic $Z^{\prime}$ bosons, 
massive color-octet states~\cite{colorOct}, Kaluza-Klein gluon excitations~\cite{gluonKK1,gluonKK2}, or 
quantum black holes~\cite{QBH1,QBH2}.

The most stringent limits on narrow, leptophobic $Z^{\prime}$ production come from the 
Tevatron.  Using 4.8~fb$^{-1}$ of data, the CDF experiment sets a 95\% C.L. limit on the mass 
of the $Z^{\prime}$ at $m_{Z^{\prime}} < 900 $~GeV/c$^2$~\cite{cdfZPrimeLJME}.  This 
analysis selects events in the lepton + jets topology and increases the signal purity by 
requiring at least four jets, at least one of which must be $b$-tagged.  The sensitivity of this 
analysis is improved using a matrix element technique in which the SM $t\bar{t}$ matrix 
element is used to construct a per event probability density function for $m_{t\bar{t}}$.  CDF 
performs a similar analysis in the all-hadronic final state with 2.8 fb$^{-1}$ of data and obtains 
a 95\% exclusion of $Z^{\prime} \to t\bar{t}$ production for $m_{Z^{\prime}} < 805$~GeV/c$^2$~\cite
{cdfZPrimeAllHadME}.  D0 excludes $Z^{\prime}$ production at the 95\% C.L. for $m_{Z^{\prime}} < 
835$ GeV/c$^2$ using 5.3 fb$^{-1}$ of data~\cite{d0ZPrimeLJ}.  In this analysis, D0 uses the lepton + 
jets for events with at least three jets, and further enhances signal purity using a neural network b-tagger.  

Currently, neither the ATLAS nor CMS experiments has the sensitivity to exclude the narrow, leptophobic 
$Z^{\prime}$ model for any range of $Z^{\prime}$ mass.  In 1.1~fb$^{-1}$ of data, CMS is able to exclude 
leptophobic $Z^{\prime}$ production where the $Z^{\prime}$ has a width $\Gamma_{Z^{\prime}} = 0.03 
M_{Z^{\prime}}$ in the mass ranges $805~\mbox{GeV/c}^2 < m_{Z^{\prime}} < 935~\mbox{GeV/c}^2$ and  
$960~\mbox{GeV/c}^2 < m_{Z^{\prime}} < 1060~\mbox{GeV/c}^2$~\cite{cmsZPrimeLJBoosted}.  This 
analysis reconstructs top pair candidates in the $\mu$+jets signature, using special reconstruction and 
event selection techniques aimed at identifying ``boosted'' $t\bar{t}$ events where the top quark decay 
products have become collimated.  Using a similar boosted reconstruction technique in 0.9 fb$^{-1}$ of 
data, CMS excludes production of Kaluza-Klein gluon excitations in a mass range of $1.0~\mbox{TeV/c}
^2 < M < 1.5~\mbox{TeV/c}^2$~\cite{cmsZPrimeAllHad}.   ATLAS searches for narrow resonances in both 
the lepton plus jets and dileptonic final states.  ATLAS excludes Kaluza-Klein gluon excitations within the 
Randall-Sundrum model for masses below 840 GeV/c$^2$ using 0.84~fb$^{-1}$ in the dilepton channel
~\cite{atlasZPrimeDIL}, and below 650~GeV/c$^2$ using 200~pb$^{-1}$ data in the lepton + jets 
channel~cite{atlasZPrimeLJ}.  Using 33~pb$^{-1}$ of data in the lepton + jets final state, ATLAS excludes 
the production of quantum black holes for black hole masses $M < 2.35$~TeV/c$^2$ at 95\% C.L.~\cite{atlasQBH}.

Beyond the narrow resonance search strategy, CDF performs a search color octet vector particles with widths ranging from $\Gamma = 0.05 M$ to $\Gamma = 0.5 M$, over a mass range from 400~GeV/c$^2$ to 800~GeV/c$^2$~\cite{cdfOctetDLM}.  In this analysis, the invariant mass distribution of events selected in the lepton + jets final state is reconstructed using dynamical likelihood method (DLM) to form an $m_{t\bar{t}}$ probability density function for each event.  The effects of interference between the color octet and Standard Model gluon are included in this search.  No signal is seen an limits are placed on the color octet coupling strength $\lambda$ as a function of the color octet particle mass and width.

Finally, to examine the $t\bar{t}$ invariant mass spectrum independent of any specific new physics model, one can measure the differential top pair production cross section as a function of the top pair invariant mass.  Using 2.7~fb$^{-1}$ of data, CDF measures $d\sigma/dm_{t\bar{t}}$ using lepton + jets events where at least one of the jets is $b$-tagged~\cite{cdfMttbarDiffXS}.  The underlying $d\sigma/dm_{t\bar{t}}$ distribution is extracted from the reconstructed $m_{t\bar{t}}$ distribution using a regularized unfolding technique based on singular value decomposition (SVD).  Consistency with the Standard Model is evaluated using the Anderson-Darling statistic with the conclusion that no evidence of new physics is present.

New high mass states can also manifest themselves as resonances in single top quark production.  The most common example of such a resonance is a heavy $W^{\prime}$ boson that arises in a number of extensions to the SM~\cite{Pati:1974yy,Mohapatra:1974hk,Senjanovic:1975rk,Mimura:2002te,Burdman:2006gy,Malkawi:1996fs,Georgi:1989xz,Perelstein:2005ka}.  Both D0 and CDF search for such new states by looking for evidence of a resonance in the reconstructed $tb$ invariant mass distribution.  Both analyses looks at data selected in the lepton + jets final state requiring a single energetic electron or muon and two or more reconstructed jets.  The CDF analysis uses a dataset corresponding to 1.9~fb$^{-1}$ and requires events to have two or three jets, with at least one jet being $b$-tagged~\cite{cdfWPrime}.  Using the reconstructed $tb$ invariant mass directly, this analysis excludes the production of a right-handed $W^{\prime}$ for $M(W^{\prime}) < 800$~GeV/c$^2$.  The D0 analysis uses 2.3~fb$^{-1}$ of data, and incorporates events with two, three, and four jets, with one or two of the jets being identified as coming from a $b$ quark~\cite{d0Wprime}.  To maximize sensitivity, each of the distinct signatures in terms of number of jets and number of $b$-tagged jets is analyzed separately.  A boosted decision tree (BDT) is used to further increase sensitivity.  The BDT incorporates input variables related to the $tb$ reconstructed mass, and kinematic properties of the individual reconstructed objects.  This analysis excludes $W^{\prime}$ masses below 863~GeV/c$^2$ to 916~GeV/c$^2$ depending on the assumptions about the allowed $W^{\prime}$ helicity and the existence of a right-handed neutrino with mass less than the $W^{\prime}$ mass.  The lowest mass exclusion $M(W^{\prime}) < 863$~GeV/c$^2$ comes in the case of a purely right-handed $W^{\prime}$ with $m(\nu_R) < M(W^{\prime})$, analogous to the scenario used for the CDF limit.  Allowing thee$W^{\prime}$ to have both left- and right-handed couplings increases the exclusion to its highest value: $M(W^{\prime}) < 916$~GeV/c$^2$.

\subsection{Forward-Backwards Asymmetry}
\label{sec:AFB}
Although a narrow resonant state with a mass larger than twice the top quark mass would most dramatically appear in the $m_{t\bar{t}}$ distribution, states with larger widths or masses beyond the reach of the accelerator in question would not be so apparent.  However, the existence of such states coupling to $t\bar{t}$ could be inferred from their effects on distributions like the top forward-backward asymmetry $A_{FB}$. Non-standard model t-channel processes can also be observed in $A_{FB}$.  For an extensive review of new physics explanations motivated by the Tevatron $A_{FB}$ results see~\cite{Kamenik:2011wt}.  Because of the differences between the $p\bar{p}$ and $pp$ initial states of the Tevatron and the LHC, there are significant differences in the approaches taken for $A_{FB}$ analyses at experiments from the two colliders.

At the Tevatron, the forward-backward asymmetry measures the relative number of events in which the direction of the top quark produced in the $t\bar{t}$ pair follows the proton direction (forward events) compared to events in which the top quark direction is oriented along the anti-proton direction (backwards events).  The precise definition depends on the reference frame used to calculate the top quark direction.  One common choice is to define the asymmetry in the $t\bar{t}$ reference frame, which is equivalent to looking at the frame-independent rapidity difference between the $t$ and $\bar{t}$ quarks $\Delta y = y_t - y_{\bar{t}}$:
\begin{equation}
A^{t\bar{t}}_{FB} = \frac{N(\Delta y > 0) - N(\Delta y < 0)}{N(\Delta y > 0) - N(\Delta y < 0)}
\end{equation}
Alternatively, one can measure the asymmetry in the laboratory frame by looking at the sign of the rapidity of the top quark measured in the lab frame:
\begin{equation}
A^{p\bar{p}}_{FB} = \frac{N(y_t > 0) - N(y_t < 0)}{N(y_t > 0) - N(y_t < 0)}
\end{equation}
Defining $A_{FB}$ in the $t\bar{t}$ reference frame has the advantage of being more directly sensitive to any $t\bar{t}$ production asymmetries, as well as being easier to interpret theoretically.  However, it also requires a complete reconstruction of the $t\bar{t}$ system, including accounting for any undetected neutrinos.  Therefore, the experimental precision of $A^{p\bar{p}}_{FB}$ is usually better than $A^{t\bar{t}}_{FB}$.

At leading order, the SM predicts no asymmetry in top pair production.  However, beyond leading order, interferences from higher-order $q\bar{q}$ annihilation diagrams result in a small asymmetry.  Calculated in the $t\bar{t}$ rest frame, this asymmetry is approximately $A^{t\bar{t}}_{FB} = 7-9\%$, while in the lab frame the asymmetry is expected to be $A^{p\bar{p}}_{FB} = 4-5\%$~\cite{AFBRef1,AFBRef2}.

Both the D0 and the CDF experiments measure $A_{FB}$ using the lepton + jets channel.  The D0 result is based on 5.4~fb$^{-1}$ of integrated luminosity~\cite{d0AFBLJ} while the CDF analysis uses 5.3~fb$^{-1}$~\cite{cdfAFBLJ}.  Both analyses require the presence of at least one $b$-tagged jet to increase the signal purity.  They also both report $A^{t\bar{t}}_{FB}$, fully corrected for detector acceptance and resolution effects.  To reconstruct the $t\bar{t}$ system, D0 and CDF use similar kinematic fitters which vary the measurements of the objects in the event according to their experimental resolutions, constraining the lepton and neutrino four-vectors as well as the hadronically-decaying $W$ boson jet four vectors to the $M_{W} = 80.4$~GeV/c$^2$, as well as the $Wb$ mass of both top quark candidates to $M_t = 172.5$~GeV/c$^2$.  The jet-to-parton assignment is resolved by chosing the combination that results in the smallest $\chi^2$ value for the kinematic fit, taking into account $b$-tagging information when considering combinations.  In the $t\bar{t}$ rest frame, D0 reports $A^{t\bar{t}}_{FB} = 0.196 \pm 0.064$ while CDF measures $A^{t\bar{t}}_{FB} = 0.158 \pm 0.074$.

Both D0 and CDF perform numerous cross checks of this result.  D0 provides the asymmetry measured in the lab frame using the charge times rapidity $q_{\ell}y_{\ell}$ from the leptonic $W$ decay to determine the rapidity of the top quark in the lab frame.  The D0 measurement $A^{\ell}_{FB} = 0.152 \pm 0.040$ can be compared to the prediction obtained using {\sc mc@nlo} for this quantity, $A^{\ell}_{FB} = 0.021 \pm 0.001$.  CDF reconstructs the asymmetry in the lab frame using the charge of the lepton from the lepton $W$ boson decay times the rapidity of the fully hadronically decaying top quark $q_{\ell}y_h$.  The CDF result $A^{p\bar{p}}_{FB} = 0.150 \pm 0.055$ can be compared to the corresponding prediction from {\sc mcfm} of $A^{p\bar{p}}_{FB} = 0.38 \pm 0.006$.  Both D0 and CDF also report asymmetries observed at various stages of event reconstruction and correction.  Of particular interest are asymmetries measured in sub-samples divided by $\Delta y$ or $M_{t\bar{t}}$ where new physics effects may be ehanced.  CDF sees a 3.4 standard deviation excess in $A_{FB}$ compared to SM expectations computed at NLO for events with $M_{t\bar{t}} \geq 450$~GeV/c$^2$.  D0 investigates the asymmetry as a function of $|\Delta y|$ and $M_{t\bar{t}}$ at the reconstructed event level (i.e. not corrected for detector acceptance and resolution) and sees no statistically significant excesses.

\begin{figure}[htbp] 
   \centering
   \includegraphics[width=0.34\textwidth]{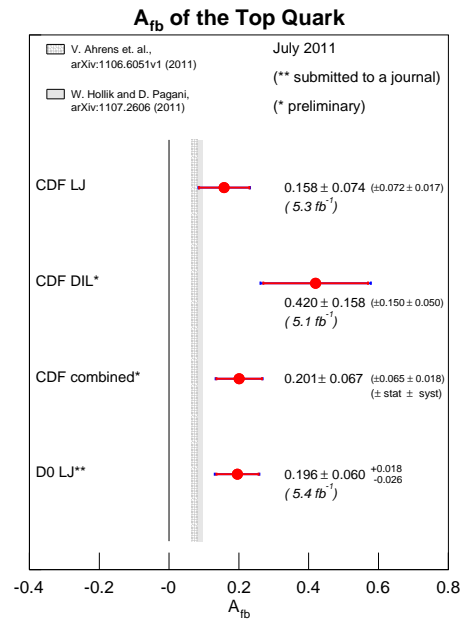} 
   \caption{Summary of CDF and D0 $A_{FB}$ measurements.}
   \label{fig:AFB}
\end{figure}

CDF also measures $A_{FB}$ for top events in the dilepton decay mode using 5.1~fb$^{-1}$ of data~\cite{cdfAFBDIL}.  Events are selected on the basis of having two energetic leptons (electrons or muons), two energetic jets, and a large amount of missing transverse energy.  This analysis does not emply $b$-tagging.  Reconstruction of the full $t\bar{t}$ system proceeds using a kinematic fitter, but the process is complicated by the presence of two undetected neutrinos.  Even applying the $W$-boson and top quark mass constraints leave the system underconstrained.  This challenge is addressed by incorporating additional constraints on $p^{t\bar{t}}_z$, $p^{t\bar{t}}_T$, and $M_{t\bar{t}}$ using probability density functions derived from SM expectations.  The result is corrected for detector acceptance, efficiency, and resolution, yielding $A^{t\bar{t}} = 0.42 \pm 0.15\mbox{ (stat)}\pm 0.05\mbox{ (syst)}$.  Combining the CDF results from the lepton + jets and dilepton channels results in a measurement of $A^{t\bar{t}} = 0.201 \pm 0.065\mbox{ (stat)}\pm 0.018\mbox{ (syst)}$~\cite{cdfAFBLJDIL}.  The Tevatron results are summarized in Fig.~\ref{fig:AFB}.

Measuring effects related to $A_{FB}$ in the $t\bar{t}$ system is particular challenging at the LHC for two reasons.  The most obvious is that the LHC has symmetric initial state.  However, there is a second issue in that non-zero $A_{FB}$ is generated--either in the SM or in new physics scenarios--via $q\bar{q}$ initiated processes which make up only a small fraction ($\sim 15\%$) of $t\bar{t}$ production at the LHC compared to the Tevatron, where $q\bar{q}$ contributes about 85\% of top pair production.  Nonetheless, some effect is visible because in $q\bar{q}$-initiated processes, the $q$ is more likely to come from a proton valence quark, while the $\bar{q}$ is pulled from the sea quarks.  There are several options for construction observables sensitive to this effect.  CMS constructs asymmetries in two ways: one based on the difference in $t$ and $\bar{t}$ quark pseudorapidities $\Delta |\eta| = |\eta_t| - |\eta_{\bar{t}}|$ and the other based on the difference of $t$ and $\bar{t}$ squared rapidities $\Delta y^2 = (y_t - t\bar{t}) \cdot (y_t + y_{\bar{t}})$.  Using data corresponding to 1.09~fb$^{-1}$ of integrated luminosity, CMS measures $A^{\eta}_C = -0.016 \pm 0.030\mbox{ (stat)} ^{+0.010}_{-0.019}\mbox{ (syst)}$ and $A^{y^2}_C = -0.013 \pm 0.026\mbox{ (stat)} ^{+0.026}_{-0.021}\mbox{ (syst)}$ consistent within uncertainties with the SM expectation~\cite{cmsCA}.  Likewise, ATLAS measures the asymmetry using a definition based on the difference in $t$ and $\bar{t}$ rapidities $\Delta y$.  In 0.70 fb$^{-1}$ of data, ATLAS obtains $A^y_C = -0.024 \pm 0.016\mbox{ (stat)} \pm 0.023\mbox{ (syst)}$, again consistent with SM expectations~\cite{atlasCA}.


\section{Summary of the current knowledge of the top quark}
\label{sec:Summary}
The Tevatron collider has now produced a sample of top quarks two orders of magnitude larger than what was needed for discovery. The top quark properties have been measured in all of the top quark final states, including the most complex ones such as the all-hadronic final states, and decays with taus. The ever-increasing dataset has allowed a rapid expansion in the horizon of feasible top quark properties measurements, pushing the experimenters ingenuity.

More than 200 published results by both the CDF and D0 collaboration establish a picture of the top quark that corresponds closely to the one predicted by the SM.   The measurement of the total cross sections for $t \bar t$ has a relative precision of 7\%, better than the current best theoretical determination; its precision can be improved analyzing further data. A better understanding of NNLO QCD computations would be needed in order to extract meaningful information. The uncertainty on the cross section for $t \bar t \gamma$ production will benefit from larger datasets. The precision on $V_{tb}$ is now of the order of 8\%; it can be reduced to 5\% using Tevatron only data. A precision beyond this value will require the larger LHC dataset, advanced understanding of the relevant systematics, and improved theoretical predictions. The single best known top quark property is the top quark mass, with a precision now approximately 0.5\%. This goal far surpassed Tevatron estimation for the Run\,II, and the goal set for the LHC collider experiments. The uncertainty on the top quark mass now translates into a negligible uncertainty on other top quark properties measurements, thus allowing stringent comparisons of the experimentally determined quantities against SM predictions. The current precision on $M_{top}$ translates into such a precision in the prediction of the Higgs boson mass that this prediction is now limited by the precision on the measurement of the $W$ boson mass. The precision on the measurement of the difference between top and antitop masses, $\approx 0.7\%$, is limited by the current statistics,  where a twofold increase in relative precision will be soon possible. The top quark width has been measured directly and indirectly, with the latter introducing more assumptions but allowing increased precision, now of the order of $30\%$. The $BR(t\to Wb)$ is measured to a 4\% level. Perhaps most interestingly, the latest measurement shows a deviation of $2.5\sigma$ from SM predictions. This measurement is already limited by the understanding of systematic sources, so an increase in statistics would not change dramatically the picture. The measurement of rare top decays are excluded up to $\approx 10^{-3}$ level. The Tevatron experiments have currently reached sensitivity to the measurement of the forward-backward asymmetry in $t \bar t$ events. This is currently the only place where the experimental determination deviates from the SM predictions by more than $3 \sigma$. More work is needed both on the theoretical and on the experimental side to understand whether this is a statistical fluctuation, an underestimation of the SM computation, or the first sign of new physics in top quark events. Measuring the same quantity for $b$ and $c$ quarks would also help to clarify the situation\,\cite{Strassler:2011vr,Bai:2011ed}.
The LHC measurements of the top charge asymmetry is in reasonable agreement with the SM prediction. Still, due to the different initial state at the LHC $p p$ collider with respect to the Tevatron $p \bar p$ collider, the two measurement provide different sensitivity to new physics scenarios.

At the time of this writing, there are only a small fraction of the measurements of top quark properties whose precision is limited by the understanding of systematics uncertainties, most notably the measurement of the top quark mass. The vast majority of the measurements are limited by the currently available datasets.


The search for new physics in samples involving top quarks has been a very active since the inception of top quark physics. 
Several searches for new physics scenarios with top quarks in the final state have been performed, without showing any sensitive deviation from the SM. The Tevatron collider expanded the LEP exclusion for 4th generation quark by about a factor of four in the exotic quark mass range. The LHC has already taken over, extending the exclusion range up to approximately 500\,GeV in mass for the up-type and down-type quarks. 
Several new physics scenarios predict top quark pairs to be produced in a resonant manner. Tevatron and LHC experiments have excluded the presence of Z' bosons, Kaluza-Klein gravitons, axigluons with masses up to approximately 1\,TeV. CDF and CMS pioneered experimental studies of the boosted top signatures as a powerful probe to extend new physics searches well beyond 1TeV. 
The LHC measurements of the top charge asymmetry will soon be able to test the SM prediction. 

\begin{table*}
\begin{center}
\caption{Cross sections for ttbar are computed by Moch and Uwer. Cross sections for single top are computed by Kidonakis.}
\begin{tabular}{lccc}
\hline\hline
Observable  & SM & Meas. & Exp.  \\
\hline
$M_{t}$ (GeV) & - & $173.2 \pm 0.9$ & Tevatron \\
$M_{t} - M_{\bar t}$\,(GeV)& 0 & $-1.2 \pm 1.3$ & CMS\\
$\Gamma_t$ (GeV) & 1.3 & $1.99^{+0.66}_{-0.55}$ & D0 \\
$Q_{top}$ & +2/3 & $\neq -4/3$ at 99\%CL & CDF \\
$V_{tb}$ & 0.998 & $0.91\pm0.08$ & Tevatron\\
$\sigma_{t \bar t}$ @\,1.96\,TeV (pb)& $7.5 \pm 0.7$ & $7.7\pm0.5$ & CDF\\
$\sigma_{t \bar t j}$ @\,1.96\,TeV  (pb)& $1.8 \pm 0.2$ & $1.6 \pm 0.5$ & CDF \\
$\sigma_{t \bar t \gamma}$ @\,1.96\,TeV (pb)& $0.17 \pm 0.03 $ & $0.18 \pm 0.08$ & CDF\\
$\sigma_{tb}$ @\,1.96\,TeV (pb) & $0.98 \pm 0.04$ & $1.8^{+0.7}_{-0.5}$ & CDF\\
$\sigma_{tq(b)}$@\,1.96\,TeV (pb)& $2.16 \pm 0.12$ & $2.90 \pm 0.59$ & D0\\
$\sigma_{t \bar t}$ @\,7\,TeV (pb)& $160 \pm 10$ & $179 \pm 12$ & ATLAS\\
$\sigma_{tq(b)}$ @\,7\,TeV (pb) &$ 65 \pm 3$ & $90^{+33}_{-22}$ & ATLAS\\
BR$(t \to Wb)$ & 0.99 & $0.90 \pm 0.04$ & D0\\
BR$(t \to Zq)$ & $\approx 10^{-12}$ & $<3.2 \times 10^{-3}$ & D0\\
BR$(t \to u g)$ & $\approx 10^{-10}$ & $<2 \times 10^{-4}$ & D0 \\
BR$(t \to c g)$ & $\approx 10^{-10}$ & $<3.9 \times 10^{-3}$ & D0\\
$F^0$& 0.698 & $0.685 \pm 0.057$  & Tev\\
$F^+$ & $<10^{-3}$ & $-0.014 \pm 0.036$ & Tev\\
$F_{SC}$ & 1 & $1.0^{+0.45}_{-0.34}$ &  ATLAS\\
$\frac{g g \to t \bar t}{q \bar q \to t \bar t}$ @ 1.96TeV & 0.15 & $0.07^{+0.14}_{-0.07}$ & CDF\\
$A_{FB}$ & 5\% & $20 \pm 7\%$ & CDF/D0 \\
$A_{FB} (m_{t \bar t} > 450GeV)$  & 8\% & $48 \pm 11\%$ & CDF \\
\hline\hline
\end{tabular}
\label{tab:nnvarQCDNN}
\end{center}
\end{table*}



\section{Future Prospects at the LHC}
\label{sec:Future}
Although the Tevatron run has concluded, the LHC is projected to have a long future.  The stated goal for the full LHC lifetime (projected to extend through 2030) is $\geq 3000$~fb$^{-1}$~\cite{myersICHEP2010}.  This projection relies on the LHC reaching instantaneous luminosities of at least $5 \times 10 ^{34}$~cm$^{-2}$s$^{-1}$.  Of course, reaching this goal will require passing through a number of intermediate phases~\cite{myersICHEP2010,myersLHCC}:
\begin{description}
\item[2012:]  As of this writing the proposal for 2012 is to increase the LHC center of mass energy to 8~TeV and to accumulate an integrated luminosity in the range of 10 - 16~fb$^{-1}$.  To achieve the high end of these projections the LHC peak instantaneous luminosity would need to increase by approximately a factor of two.
\item[$\sim$2015-2020:]  After a shutdown lasting between one and two years to repair the magnet splices, the LHC would begin running again at center of mass energy somewhere between 13~TeV and 14~TeV.  Planning to reach peak instantaneous luminosities up to $2 \times 10^{34}$~cm$^{-2}$s$^{-1}$, the LHC should deliver an integrated luminosity starting around 20-30~fb$^{-1}$ in the first year, and eventually reaching as high as $\sim 100$~fb$^{-1}$.
\item[$\sim$2021-2030:]  After another shutdown for upgrades to increase the LHC luminosity, as well as to improve the detectors ability to handle high pileup environments, the LHC will begin running in a new high luminosity regime.  In this era, peak instantaneous luminosities would reach as high as $5 \times 10^{34}$ cm$^{-2}$s$^{-1}$, corresponding to on the order of 100 pileup $pp$ interactions per bunch crossing.  Annual integrated luminosity totals would range between 100 fb$^{-1}$ to 300 fb$^{-1}$.
\end{description}

The tremendous growth in the top quark samples at the LHC will allow measurements of top quark properties with extreme precision, effectively removing the statistical uncertainty many measurements.  Cross section measurements will evolve from measurements of inclusive production to measurements of differential and double-differential distributions.  The large datasets and higher energies will also greatly extend the reach of searches, such as the search for a heavy resonance decaying to top quark pairs or additional generations of quarks with top-like signatures.  Measurements of objects produced in association with top quark pairs, such as $t\bar{t}Z$, $t\bar{t}W$, $t\bar{t}b\bar{b}$, and $t\bar{t}t\bar{t}$, will reach SM level of sensitivity, probing new physics scenarios that would show up as enhancements in these rates.  Depending on how the search for the Higgs boson unfolds, one of the potentially most interesting processes connected to the top quark, production of top quarks in association with a Higgs boson, could begin to become accessible when the dataset reaches the level of 30-100 fb$^{-1}$ of integrated luminosity~\cite{atlasTDR2,atlasTDR1,cmsTDR2,cmsTDR1}.  Finally, increasingly rare top quark decays will become accessible, from BSM decays like $t \to H^+b$ to the SM decay $t \to WZb$.

An important consideration in all of these measurements is that most of them will rapidly become limited by systematic uncertainties, particularly those related to background predictions, unless further progress can be made to improve them beyond current levels.  For example, studies suggest that theoretical uncertainties on the irreducible $t\bar{t}b\bar{b}$ background will significantly reduce the significance of the $t\bar{t}H$ signal unless the uncertainties can be reduced from present levels.  Therefore, it is quite likely that further progress in top quark physics will follow in the wake of improvements in modeling the background fueled by insights derived from the high statistics LHC dataset.

Finally, despite the high statistics, some measurements from the Tevatron will remain difficult to repeat at the LHC.  In particular, any measurement relying on $q\bar{q} \to t\bar{t}$ production will be difficult to access among the high rate of $gg \to t\bar{t}$.  For example, as mentioned previously, the intriguing excess in $A_{FB}$ observed at the Tevatron will not be directly accessible at the LHC, and alternative strategies will have to be employed to probe effects related to $A_{FB}$ at the LHC.

\section{Conclusions}
\label{sec:Conclusions}
The Tevatron collider just ended its second and last Run, producing two orders of magnitude more top quark events than its Run\,I predecessor. This large increase allowed a big leap forward in the understanding of the top quark. All $t \bar t$ final states have now been unambiguously observed ($5\sigma$); the electroweak single top quark production has been observed in the final state including leptons, providing the first direct measurement of $V_{tb}$. Many analysis techniques and tools useful for top quark physics have been tested and refined. The top quark mass has been measured with very large precision ($<1\%$) and the experimental precision of the top quark pair production cross section now surpasses the theoretical one. The study of other top quark properties, as well as the direct search for new physics involving top quarks has largely advanced, but still they are mostly statistically limited. The top quark as of today appears to be mostly the one predicted by the SM, with the notable exception of the anomalously large forward-backward asymmetry, showing multiple deviation at the $3\sigma$ level from the NLO prediction, and confirmed by both Tevatron experiments. This is the most compelling example of the uniqueness of the $p \bar p$ collisions at the Tevatron.  As most measurements---including the aforementioned---use only half of the existing dataset, it will be crucial to obtain the final Tevatron results. 

The LHC collider restarted its operation in 2009 after a the shutdown induced by faulty electrical connection between the acceleratorÕs magnets; in summer 2011 the number of collisions producing top quarks equaled the Tevatron sample. Thanks to the spectacularl LHC performances, at the time of this writing the ATLAS and CMS top quark datasets already exceed Tevatron's by one order of magnitude.  Several measurements already surpassed  in precision previous Tevatron results, and tools for the exploration of NP beyond the TeV scale in top events have been tested successfully. With the ever-increasing dataset, the search for rare top decays will become increasingly important, as well as the production of top quarks together with heavy vector bosons ($W/Z$), and eventually with the Higgs. All of the above measurements would test stringently the SM, while any deviation would be a clear sign of new physics. Direct searches for new physics will allow soon the observation or exclusion conclusively of the existence of $4^{\mathrm{th}}$ generation quarks, as well as the existence of SUSY scalar partners of top quarks. 

This is the most interesting time ever for top quark physics: the multitude of SM tests in top quark properties measurements, its increasing connection to flavor physics, and the fact that the top quark sits at the highest SM energy scale, guarantee that the heaviest SM particle will be captivating the attention of physicists for years to come.

\bibliographystyle{epjc}
\bibliography{kevin,fabrizio,neu_refs}
\end{document}